\documentclass{aa}  

\usepackage{ulem}
\usepackage{comment}
\usepackage{graphicx}
\usepackage[dvipsnames]{xcolor}
\usepackage{caption}
\usepackage{subcaption}

\usepackage{pdflscape} % Added by Marcel
\usepackage{txfonts}
\usepackage{hyperref} 
\hypersetup{colorlinks=true, linktocpage=true, pdfstartpage=3,
            pdfstartview=FitV, breaklinks=true, pdfpagemode=UseNone,
            pageanchor=true, pdfpagemode=UseOutlines, plainpages=false,
            bookmarksnumbered, bookmarksopen=true, bookmarksopenlevel=1,
            hypertexnames=true, urlcolor=RoyalBlue,
            linkcolor=RoyalBlue, citecolor=RoyalBlue}

\usepackage[utf8]{inputenc}
\usepackage[T1]{fontenc}
            
\newcommand{\rem}[1]{}

\begin{document}

   \title{Hubble Asteroid Hunter III. \\ Physical properties of newly found asteroids}

   %\subtitle{}

   \author{Pablo Garc\'ia-Mart\'in\inst{1},
          Sandor Kruk \inst{2}, Marcel Popescu \inst{3,4}, Bruno Mer\'in  \inst{2}, Karl R. Stapelfeldt \inst{5}, Robin W. Evans \inst{6},  Benoit Carry \inst{7}, Ross Thomson \inst{8}
          }

   \institute{Department of Theoretical Physics, Autonomous University of Madrid (UAM), Madrid 28049, Spain
   \and
   European Space Agency (ESA), European Space Astronomy Centre (ESAC), Camino Bajo del Castillo s/n, 28692 Villanueva de la Ca\~nada, Madrid, Spain
  \and
  Astronomical Institute of the Romanian Academy, 5 Cutitul de Argint, 040557 Bucharest, Romania
  \and
  University of Craiova, Str. A. I. Cuza nr. 13, 200585 Craiova, Romania
  \and
    Jet Propulsion Laboratory, California Institute of Technology, Mail Stop 321-100, 4800 Oak Grove Drive, Pasadena, California 91109, United States
  \and
  Bastion Technologies, 17625 El Camino Real 330, Houston, TX 77058, United States
  \and
  Universit\'e C\^ote d’Azur, Observatoire de la C\^ote d’Azur, CNRS, Laboratoire Lagrange, France
  \and
  Google Cloud, 6425 Penn Ave, Pittsburgh, PA 15206, United States
}

   \date{Received April 28, 2023; accepted December 22, 2023}

\titlerunning{Hubble Asteroid Hunter III. Physical properties of newly found asteroids}
\authorrunning{Garc\'ia Mart\'in, et al.}
% \abstract{}{}{}{}{} 
% 5 {} token are mandatory
 
  \abstract
   {Determining the size distribution of asteroids is key for understanding the collisional history and evolution of the inner Solar System.}
   {We aim at improving our knowledge on the size distribution of small asteroids in the Main Belt by determining the parallaxes of newly detected asteroids in the Hubble Space Telescope (HST) Archive and hence their absolute magnitudes and sizes. }  
   {Asteroids appear as curved trails in HST images due to the parallax induced by the fast orbital motion of the spacecraft. Taking into account its trajectory, the parallax effect can be computed to obtain the distance to the asteroids by fitting simulated trajectories to the observed trails. Using distance, we can obtain the object’s absolute magnitude and size estimation assuming an albedo value, along with some boundaries for its orbital parameters.}
   {In this work we analyse a set of 632 serendipitously imaged asteroids found in the ESA HST Archive. These objects were obtained from instruments ACS/WFC and WFC3/UVIS. An object-detection machine learning algorithm (trained with the results of a citizen science project) was used to perform this task during previous work. Our raw data consists of 1,031 asteroid trails from unknown objects, not matching any entries in the Minor Planet Center (MPC) database using their coordinates and imaging time. We also found 670 trails from known objects (objects featuring matching entries in the MPC). After an accuracy assessment and filtering process, our analysed HST asteroid set consists of 454 unknown objects and 178 known objects. We obtain a sample dominated by potential Main Belt objects featuring absolute magnitudes (H) mostly between 15 and 22 mag. The absolute magnitude cumulative distribution $log N(H>H_0)\propto\alpha\log(H_0)$ confirms the previously reported slope change for $15<H<18$, from $\alpha \approx 0.56$ to $\alpha \approx 0.26$, maintained in our case down to absolute magnitudes around $H\approx 20$, hence expanding the previous result by approximately two magnitudes. }
   {HST archival observations can be used as an asteroid survey since the telescope pointings are statistically randomly oriented in the sky and they cover long periods of time. They allow to expand the current best samples of astronomical objects at no extra cost on telescope time.}


   \maketitle
%
%-------------------------------------------------------------------
\section{Introduction}
\label{introduction}

The Solar System objects (SSOs) can be studied by observing their movement over time, which requires multiple exposures separated by a few days to produce a sufficiently large orbital arc. If the SSOs have been previously reported to the Minor Planet Center (MPC), the main worldwide repository for the receipt and distribution of positional measurements of SSOs\footnote{\url{https://www.minorplanetcenter.net/}}, precovery in archival data can aid in orbit determination. In this paper, we use a different method for characterising asteroids using a set of observations taken from low Earth orbit in large astronomical archives.

\begin{figure}
   \centering
   \includegraphics[width=\columnwidth]{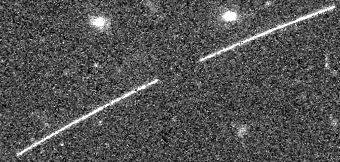}
   \caption{Example of asteroid trail imaged by HST in observation idj554030 from WFC3/UVIS, taken on Feb. 5, 2018. The trail spans roughly 10 arcsec in Right Ascension and 8 arcsec in Declination.}
    \label{trail_1st_example}%
\end{figure}

Observatories in low Earth orbit have a distinct advantage, as their fast motion produces a parallax effect on the asteroid's trail shape, allowing for distance determination with fewer observations. This parallax method was first devised by \citet{Evans1998, Evans2002} for use with asteroids identified in Hubble Space Telescope (HST) observations. An example of an asteroid trail, imaged by HST, where the parallax effect is evident in the curvature of the trail is shown in Figure \ref{trail_1st_example}.

In the Hubble Asteroid Hunter project\footnote{\url{www.asteroidhunter.org}} \citep{hah1}, we identified 1701 new asteroid trails in 19 years of HST observations with the Advanced Camera for Surveys(ACS/WFC) and Wide Field Camera 3 (WFC3/UVIS) observations, using a deep learning algorithm on Google Cloud, trained on volunteer classifications from the Hubble Asteroid Hunter citizen science project on the Zooniverse platform. 1031 (61\%) of the trails correspond to SSOs that we could not identify in the MPC database. Considering how faint these asteroids are, many of them are likely previously unidentified objects. The parallax fitting method \citep{Evans1998} is ideal to study this new sample of SSOs. By fitting the parallax for these trails, we can determine the approximate orbital parameters of these asteroids and derive their physical properties. Once we determine the distance to the asteroids and using their apparent magnitudes, we can determine their intrinsic brightnesses and sizes, assuming an average albedo value.

Studying faint asteroids is essential for better understanding their size distribution, which remains poorly understood for small objects. The shape of the magnitude distribution curve, as determined by \citet{Gladman2009}, shows a shallower slope for asteroids of smaller sizes down to $H=18$ mag. This result is crucial to understand the Main Belt evolution along with its collisional and dynamic depletion models \citep{Bottke_ast4}.

Hubble Space Telescope (HST) observations, taken over a long time span, can be effectively used as a randomly observed asteroid survey since the telescope's pointing selections are statistically random in the sky, and many observations are across the ecliptic where the Main Asteroid Belt is located.

This paper is structured as follows. In Section 2, we describe the data that was utilized in this study. In Section 3, we outline the parallax fitting method and examine its efficacy through tests conducted on both simulated data and a sample of known asteroids in the MPC database, also identified by the Hubble Asteroid Hunter citizen science project volunteers. Section 4 reports the results of the application of this method to previously unidentified asteroids, including the determination of their orbital parameters and size distributions. In Section 5, we compare our findings to those of previous studies. Finally, in Section 6 we present our conclusions.

\section{Data}
\label{datasection}
The trails analysed in this work were obtained from the ESA Hubble Space Telescope Archive for instruments ACS/WFC and WFC3/UVIS. An object-detection machine learning algorithm (trained with the results of a citizen science project) was used to find asteroid trails \citep{hah1} and other objects of interest such as gravitational lenses \citep{hah2} and satellite trails \citep{natastron_sats} in Hubble images from the archive for both instruments. We used a total of 24,731 HST composite images from ACS/WFC and 12,592 from WFC3/UVIS, each one of them split in four quadrant cutouts for improving asteroid trail detection efficiency. HST images from the archive were selected based on the following criteria: no calibration images or 'darks', exposure time greater than 100 seconds, FoV greater than 0.044$^o$ (avoiding sub- frames), and no grism spectral images, as their special features could be confused with trails. We also removed any image purposely targeting Solar System objects.

The volunteers of our citizen science project and also the machine learning algorithm were trained to identify cosmic rays to correctly distinguish their streaks from real asteroids. Nevertheless, the whole set of images featuring potential detections was visually checked by the authors during the first part of this project to avoid any artifact or spurious detection misidentified as an asteroid.

Our data consists of 1,031 asteroid trails from unknown objects, not matching any entries in the MPC database based on their coordinates and imaging time. We also found 670 trails from known objects (objects featuring matching entries in the MPC database), which will prove very useful to test and validate the method we propose in this work. For every case, a list of right ascension and declination coordinates forming a trail is used along with the exposure starting and ending times.

Checking the identified trails for known objects was conducted using a two step process. The first step involved using the average position and exposure time of each trail. We fed this data to the SkyBoT service \citep{Skybot1, Skybot2} performing a cone search with a 900 arcsec radius. This preliminary step returned 22,748 possible candidates for our whole set. During the second step we filtered these candidates for precise matches with the observed trails. For every image, ephemeris for all its possible matches was retrieved from JPL Horizons service \citep{jpl_horizons} using the exposure time. Then we analysed the approximate averaged distance between the observed trail and those obtained from JPL Horizons (as seen from HST), using a 3 arsec threshold for a positive match. The complete trail processing pipeline applied during the first part of this project is explained in full detail in \citet{hah1}.

\section{Analysis methods}
\label{methods}

\subsection{Parallax method}
\label{parallax_section}

To obtain the distance to Earth of the unknown objects, we present a method using the parallax-induced curvature of asteroid trails in HST images, following the work by \citet{Evans1998} and \citet{Evans2002}. Asteroid trails imaged by HST present a very particular parallax induced by the spacecraft orbital motion around Earth while tracking a fixed target. Being in a low Earth orbit, HST's fast pace compared to its typical exposure times enables this unique effect for relatively "close" objects in our Solar System. During a 30-minutes exposure (the average time for our dataset) HST travels approximately a third of its orbit, a stationary object situated at 2 AU from Earth (around the Main Belt) could feature a parallax effect of up to $\approx$ 8 arcsec.

For every trail, HST orbital motion for the exposure time is obtained from Jet Propulsion Laboratory (JPL) SPICE routines \citep{spice,spice2} along with Earth position with respect to the Solar System Barycenter (SSB). Our code iterates through different object distance solutions between 0.2 and 6.7 AU, obtaining predicted trails which are compared to the observed trail looking for a best-fit solution. The 6.7 AU upper distance limit was chosen to potentially cover Jupiter Trojan asteroids population and also based on the non-curvature of trails for objects located further.

In addition to the parallax induced by the object's distance and HST-Earth trajectories, we also need to consider the object's motion around the Sun, which we will call the \textit{object's intrinsic rate of motion} as it is observed in our images.

The observed trail is the linear combination of the parallax effect modelled by vector $P(\Delta, t)$, a function of both observer distance (\textit{$\Delta$}) and time (\textit{t}); and the \textit{object's intrinsic rate of motion} in right ascension and declination: \textit{$\frac{d\alpha}{dt}$} and \textit{$\frac{d\delta}{dt}$} respectively.

From \citet{Evans1998}, every point of the trail can be then described as:
\begin{equation}
\begin{aligned}
\alpha - \alpha_{0} = P_{\alpha}(\Delta,t) + \frac{d\alpha}{dt}\cdot t\\
\delta - \delta_{0} = P_{\delta}(\Delta,t) + \frac{d\delta}{dt}\cdot t
\label{eq:parallax}
\end{aligned}
\end{equation}

\noindent{where \textit{$\alpha_0$} and \textit{$\delta_0$} are the initial coordinates of the object; \textit{$P_{\alpha}$} and \textit{$P_{\delta}$} are the components of vector \textit{P}; \textit{$\frac{d\alpha}{dt}$} and \textit{$\frac{d\delta}{dt}$} are the object' intrinsic rates of motion and \textit{t} is the time of the exposure being considered.}
\\

To obtain the observed \textit{object's intrinsic rate of motion} we use a rather straightforward method. The total span of the trail in both right ascension and declination coordinates must be equal to the asteroid motion rate plus the "deformation" generated by the parallax effect. Knowing the starting and ending points of the trail and the standard "deformation" generated by the parallax at any given distance, just one unique value of \textit{$\frac{d\alpha}{dt}$} and \textit{$\frac{d\delta}{dt}$} is possible for each distance iteration.

The selected trail solution is the one featuring the lowest error value ${\chi}^2$ as defined in equation \ref{eq:chisquare}, comparing the predicted trails for 1,300 different observer distance values ($\Delta$) and the observed trail. We normalise ${\chi}^2$ using a 0.05 arcsec value, which corresponds to the ACS/WFC pixel resolution \citep{ACShandbook}. WFC3/UVIS resolution is 0.04 arcsec/pixel \citep{WFChandbook}, but we decided to use the worst ACS/WFC case for our analysis to present an unified method for both instruments.

The ${\chi}2$ expression is as follows:

\begin{equation}
\begin{aligned}
\chi^2 =\frac{1}{n} \sum \left(\frac{x_{calc}-x_{obs}}{0.05} \right)^2
\label{eq:chisquare}
\end{aligned}
\end{equation}
\noindent{where \textit{n} is the total number of existing points for the trail, \textit{$x_{calc}$} are the sky position values from the simulated trail being fitted and \textit{$x_{obs}$} are the observed trail sky position values.}
\\

During the first stage of this project \citep{hah1} the trails were discretised by our extraction algorithm finding the maxima along each pixel column of the image cutout file. Taking into account HST resolution, this is approximately one point every 0.04 arcsec for WCS3/UVIS and 0.05 arcsec for ACS/WFC. The average number of points for a given trail is $\approx$ 300.

 We cannot distinguish between distance solutions with a mean fitting-residual difference lower than the instrument resolution (considered to be one pixel, 0.05\arcsec). Therefore the uncertainty $U_r$ of our method is set by the two distance values corresponding to ${\chi}^2+1$ (as shown in Figure \ref{good_ast_chisq}).

Once we have a distance solution for a trail as seen from HST, we obtain the distance to Earth and SSB using SPICE routines. Given that all the analysed asteroids were serendipitously imaged during a single HST observation, we cannot apply any standard orbit approximation method. The object's velocity component in the asteroid-Earth direction will always be missing and we can only obtain the limits of its possible orbital parameters for semi-major axis (a), eccentricity (e) and inclination (i).

Nevertheless, we can calculate some possible combinations of these parameters supposing a bounded orbit. The object's estimated velocity and escape velocity are calculated from the observation (the distance from the Sun can be easily obtained from Earth distance and time using SPICE) and 20,000 different velocity values for the missing asteroid-Earth axis velocity component are iterated using the escape velocity as a limit. We obtain a set of 20,000 possible orbital parameters combinations for each object, which can be plotted as a function of semi-major axis as seen in Figure \ref{good_ast_orbitparams}. We can also obtain minimum semi-major axis, eccentricity and inclination values for the object. Objects featuring estimated velocities greater than their escape velocity are not considered in our analysis and are expected to be cases where the method will not be able to obtain a good result (see Section \ref{accuracy} for further details on the filtering process).

\begin{figure}
   \centering
   \includegraphics[width=\columnwidth]{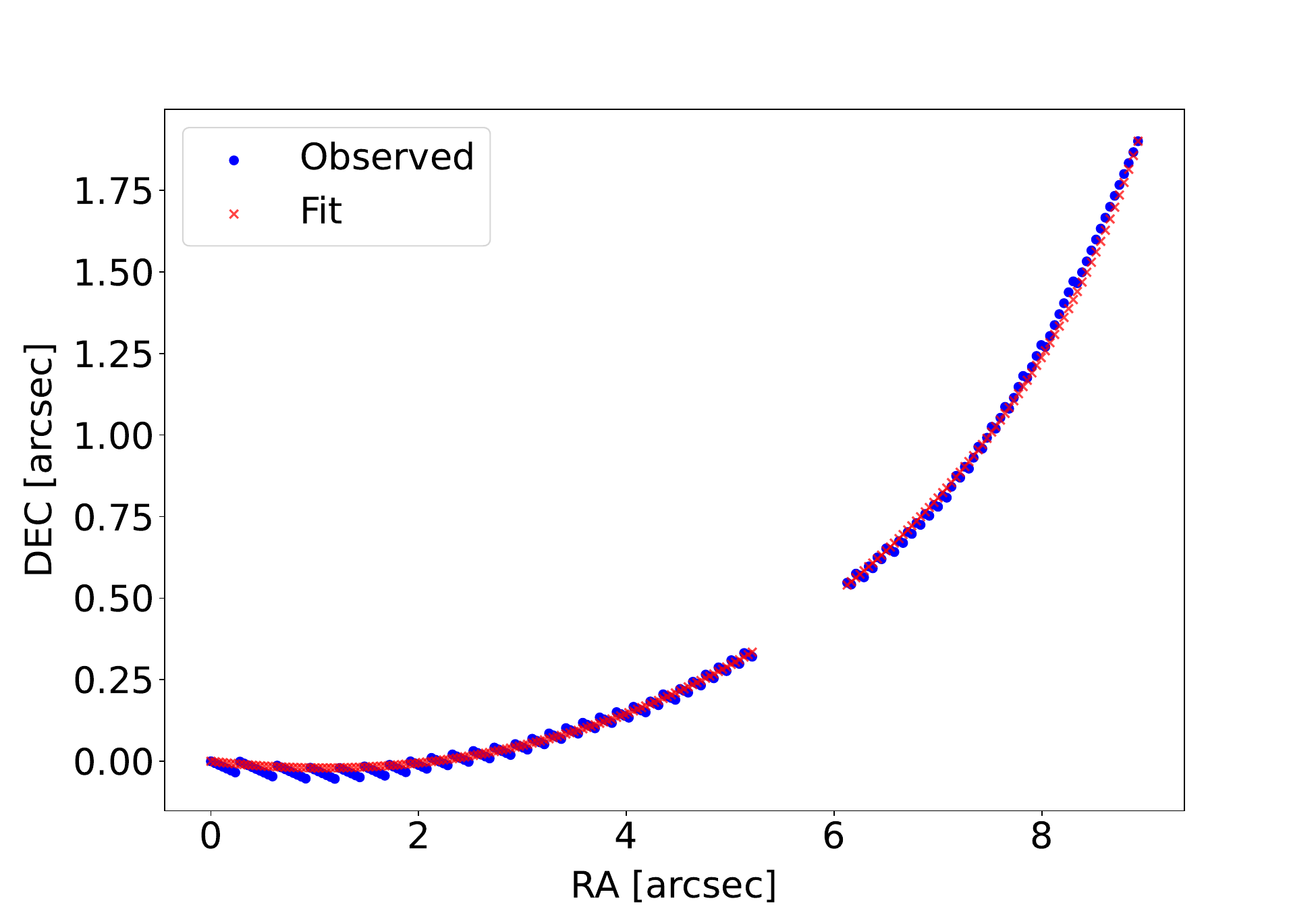}
   \caption{Asteroid (511908) 2015 HU67. Observed trail from HST (blue) vs best-fit distance solution using the parallax method (red). ${\chi}^2$ value for this fitting is 0.13 (see equation \ref{eq:chisquare}).}
    \label{good_ast_fit}%
\end{figure}

\begin{figure}
   \centering
   \includegraphics[width=\columnwidth]{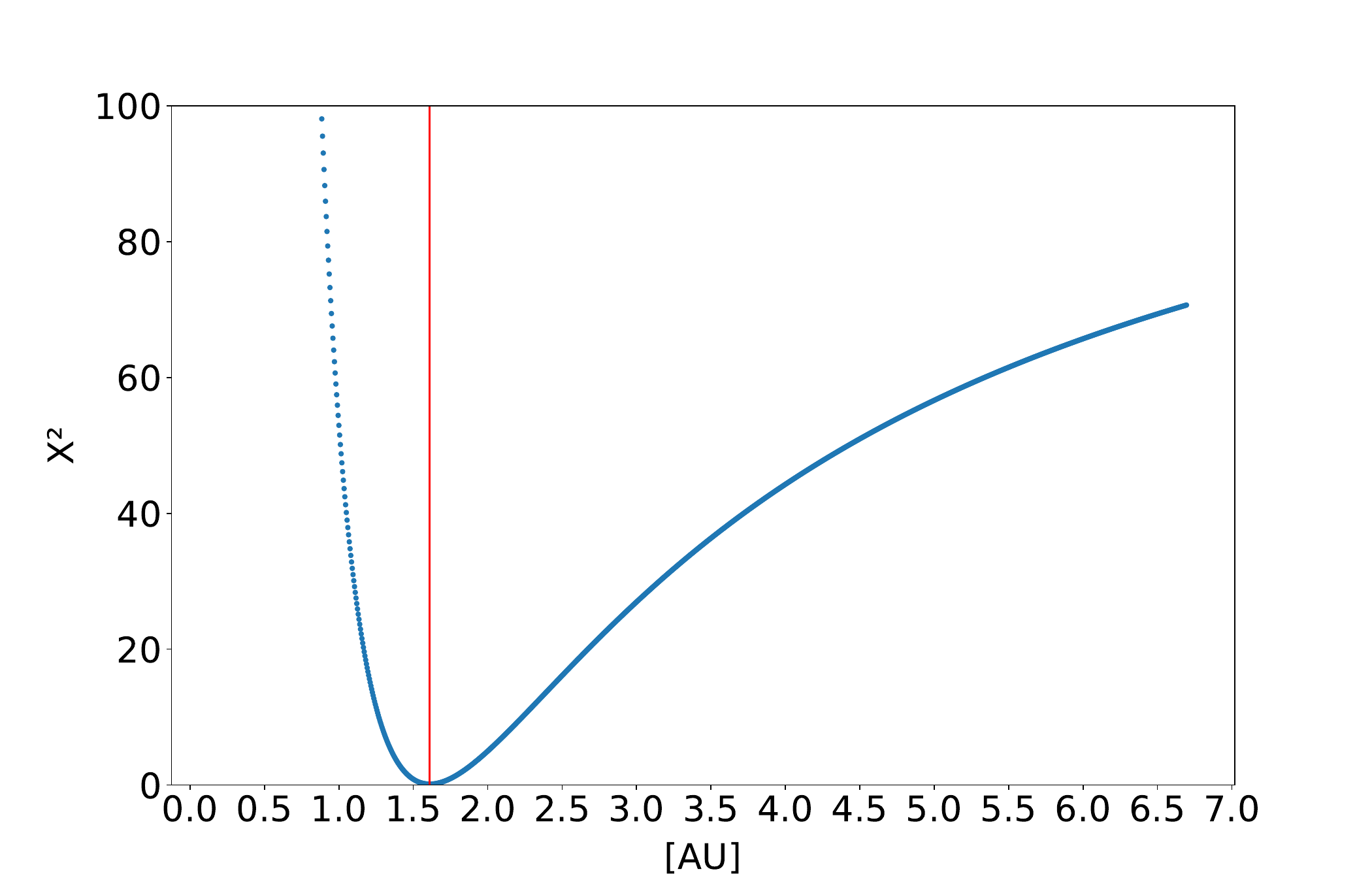}
   \caption{Asteroid (511908) 2015 HU67. ${\chi}^2$ values showing a minimum at the best-fit distance solution from Earth. The vertical line represents the actual distance for this object obtained from JPL Horizons ephemeris.}
    \label{good_ast_chisq}%
\end{figure}

\begin{figure}
   \centering
   \includegraphics[width=\columnwidth]{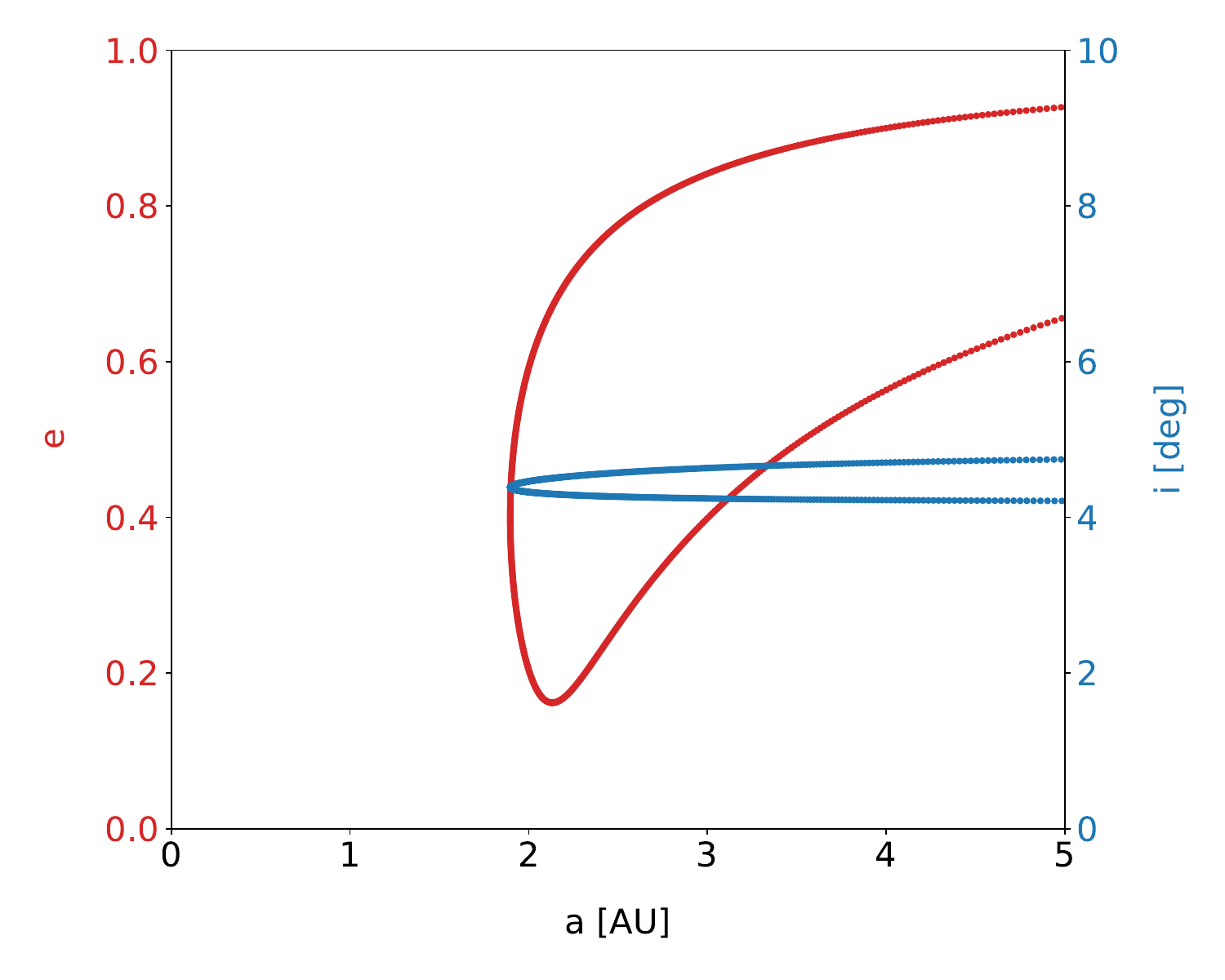}
   \caption{Possible values of eccentricity (e, red) and inclination (i, blue) as a function of semi major axis (a) for asteroid (511908) 2015 HU67 calculated using the parallax  method. For every semi major axis value, the corresponding possible eccentricity and inclination solutions are shown graphically}
    \label{good_ast_orbitparams}%
\end{figure}

As an example of the parallax method, we show the results for object (511908) 2015 HU67 identified in HST observation idq27j020. This object is a relatively small asteroid from the middle Main Belt featuring an apparent magnitude of 21.5 in the V band. This trail was processed using our full code pipeline, from trail coordinates to orbital parameters results. The trail fit and ${\chi}^2$ minimum are shown in Figure \ref{good_ast_fit} and Figure \ref{good_ast_chisq} respectively. The graph featuring all possible inclination and eccentricity values as a function of semi-major axis is shown in Figure \ref{good_ast_orbitparams}.

As a benchmark, we use JPL Horizons ephemeris service \citep{jpl_horizons} to validate the parallax-obtained Earth distance and orbital parameters for this known object. We show this comparison in Table \ref{good_ast_results}. Our parallax results show a good agreement with the values obtained from the database. 

In some cases the trail does not present enough curvature to converge to a well defined distance solution or to obtain a reduced uncertainty interval allowing us to calculate a meaningful absolute magnitude value. If the trail is not curved enough, our ${\chi}^2$ function will be too broad, implying a large uncertainty $U_r$ in the distance solutions. Trails lacking significant curvature correspond to objects seen close to HST's orbital plane, i.e. the induced parallax is close to the direction of the asteroid's intrinsic motion. The most curved trails are those where the asteroid is seen at a large angle from HST's orbital plane, such that the induced parallax is becoming perpendicular to the asteroid's intrinsic motion.

\begin{figure}
   \centering
   \includegraphics[width=\columnwidth]{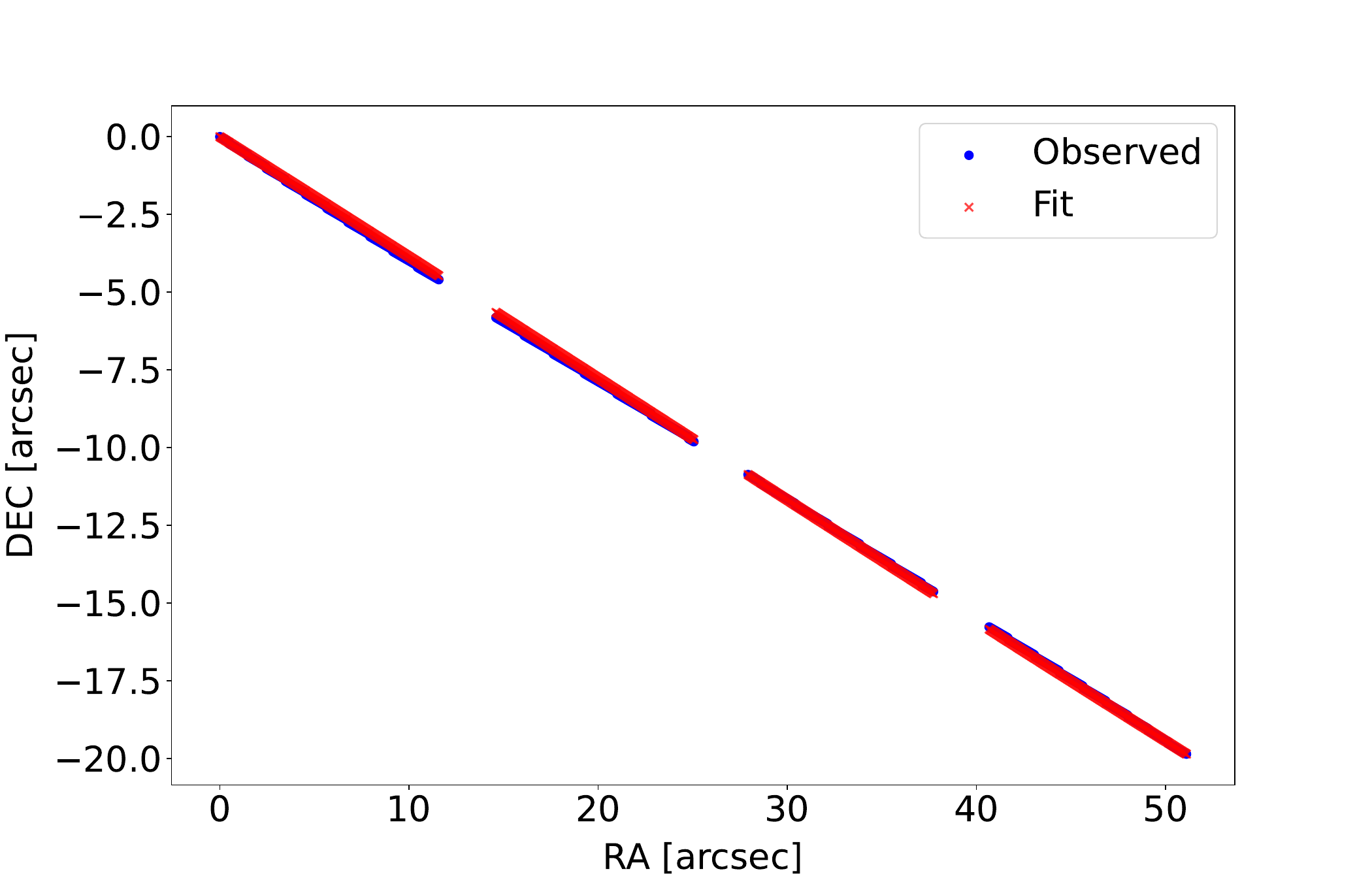}
   \caption{Object (58250) 1993 QU1. Example of straight trail, with no parallax solution obtained.}
    \label{bad_ast_fit}%
\end{figure}

\begin{figure}
   \centering
   \includegraphics[width=\columnwidth]{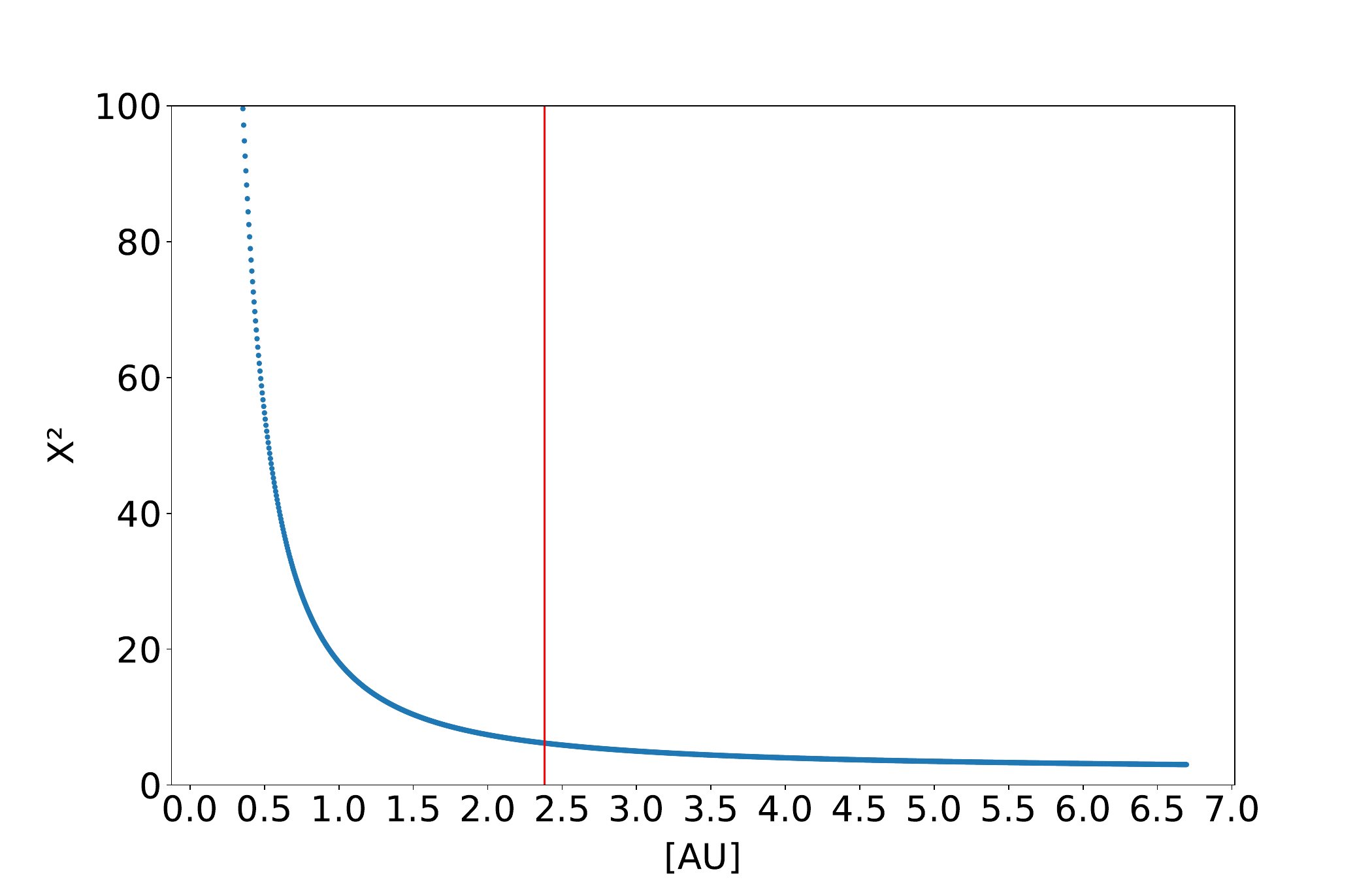}
   \caption{Object (58250) 1993 QU1. The method does not converge because of the lack of curvature from the trail. The ${\chi}^2$ minimum is obtained at the upper distance limit set in our code ($\approx$ 6.7 AU). The red line represents the actual Earth distance for this object from JPL Horizons ephemeris.}
    \label{bad_ast_chisq}%
\end{figure}

An example of a trail not presenting a clear solution is object (58250) 1993 QU1 featured in HST image with observation id j93y94010. Despite its high signal to noise ratio and good fit (Figure \ref{bad_ast_fit}), the trail is too straight to yield an accurate distance solution and our method does not converge (Figure \ref{bad_ast_chisq}). We can also see in Figure \ref{bad_ast_chisq} that the trail fitting yields a large number of possible solutions inside the ${\chi}^2+1$ threshold, as expected for straight trails. The obtained Earth distance value is 6.695 AU (the actual JPL Horizons Earth distance equals 2.382 AU). These cases were filtered out from our sample as discussed in Section \ref{accuracy}. 

\begin{table}[]
    \centering
    \caption{Orbital parameters of (511908) 2015 HU67 from parallax method vs JPL Horizons ephemeris data.}
    \begin{tabular}{  c | c | c  } 
    \hline\hline
    Feature & Parallax results & JPL Horizons\\ 
    \hline
    Earth dist. [AU] & 1.610 & 1.607 \\ 
    % \hline
    SSB dist. [AU] & 2.406 & 2.404 \\
    % \hline
    Phase Ang. [deg] & 17.941 & 18.137 \\
    % \hline
    Ecliptic lat. [deg] & 1.824 & 1.820 \\
    % \hline
    e & 0.162 < e & 0.275 \\
    % \hline
    i [deg] &  4.178 < i < 4.909 & 4.733\\
    % \hline
    a [AU] & 1.897 < a & 2.585\\
    \hline
    \end{tabular}
    \label{good_ast_results}
\end{table}

\subsection{Assessing the accuracy of the method}
\label{accuracy}

\begin{figure}
   \centering
   \includegraphics[width=\columnwidth]{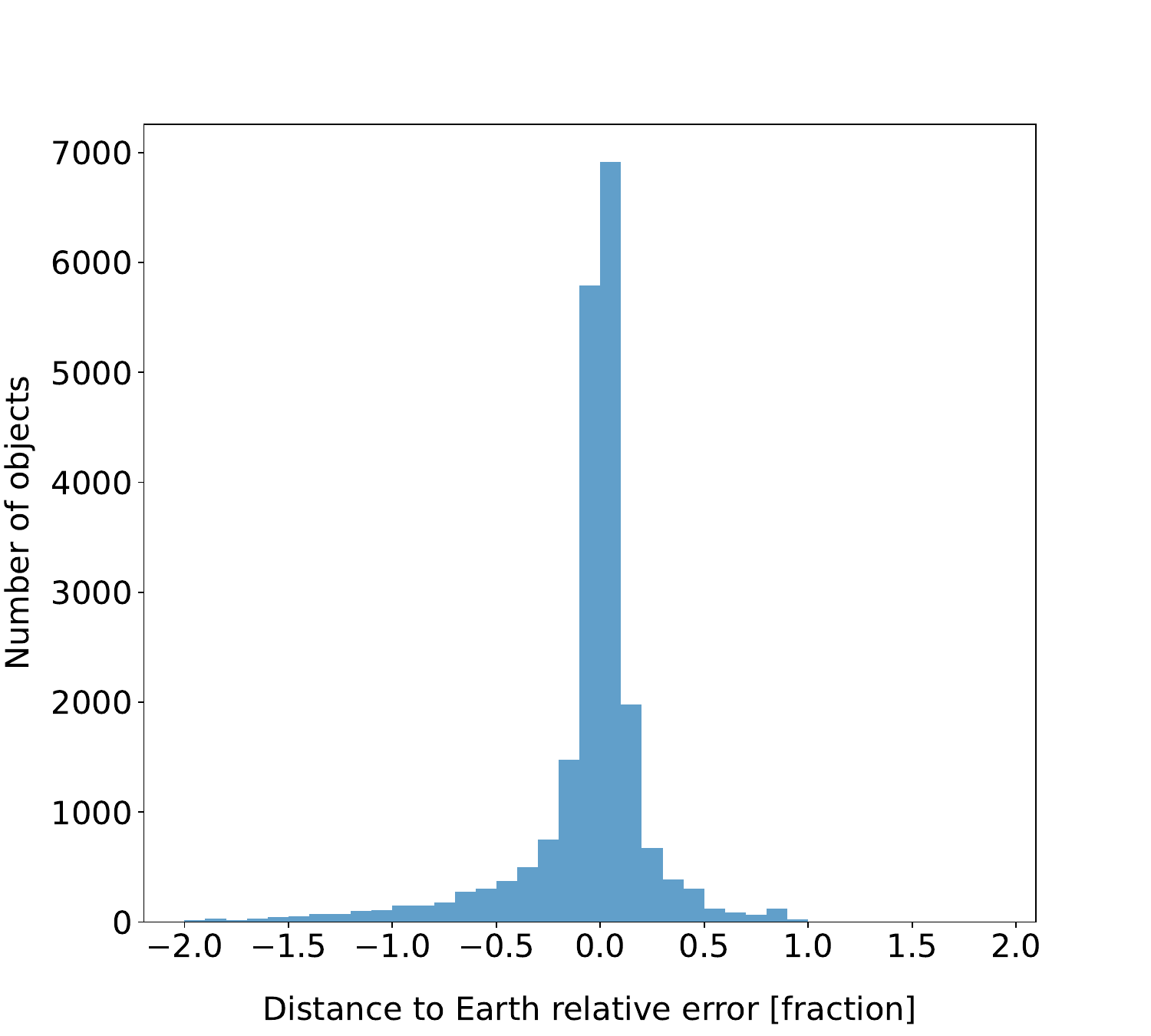}
   \caption{Relative distance error for JPL Horizons ephemeris distance vs parallax-obtained distance for the proof of concept set (21,283 known objects) using coordinates from JPL Horizons as simulated trails (RA, DEC and time). x axis is $ distance \ error/actual \ distance $. No filtering or goodness of fit criteria was applied to this proof of concept, which explains the long tails at both ends of the distribution.}
    \label{JPL_histogram}%
\end{figure}

To catch a first glimpse of the the accuracy of our method, we used 21,283 available known-asteroid ephemeris from JPL Horizons service \citep{jpl_horizons} as a "proof of concept" (POC) set. For every object, we input only the point coordinates (RA and DEC) and time, as a real trail observed by HST, and applied our method to calculate a distance solution. The obtained distance from Earth was then compared with the distance value provided from JPL Horizons service. The global relative distance error $(distance \ error / real\ object\ distance)$ results for this POC set are centered around 0 and can be seen in Figure \ref{JPL_histogram}. This confirms that our parallax method produces distance results compatible with JPL Horizons database values. No filter or goodness of fit criteria was applied to the results of this analysis or to the input data, which explains the long tails at both ends of the error distribution. This result validates the accuracy of our Python implementation of the method from \citet{Evans1998} which was originally implemented in Fortran.

The next step is to evaluate our method in more detail and precisely define its accuracy for distance predictions. To do so we used the 670 trails from known objects identified in the first part of this project \citep{hah1}. Out of these 670 asteroids, we selected the 452 featuring a permanent Minor Planet Center (MPC) designation to reduce potential errors in their reference orbital parameters. We ran the code using this group, which we name known-objects-validation set, and after examination we decided to apply the following criteria to filter spurious distance results that could contaminate our final population distribution. We dropped:

\begin{itemize}
  \item Objects presenting observed estimated velocities above their escape velocity. No extra-solar objects were included in the known objects set and the probability of finding one among our results was deemed extremely unlikely.
  \item Objects above a 6.5 AU distance result or not featuring a clear ${\chi}^2$ absolute minimum ($U_r$ uncertainty equal to 0), for which the method was not converging.
  \item Objects featuring an upper or lower distance uncertainty $U_r$ value equal to or greater than 35\% of their actual distance. We chose this value analysing the relative error distribution for the whole raw set ($ distance \ error/actual \ distance $). It represents one standard deviation ($\sigma$) for this distribution. As mentioned before, these last two criteria are meant to filter trails too straight to yield a reliable parallax solution.
  \item Trails as non-monotonic functions (RA or Dec acting as X-Y, see an example in Fig 17 of \cite{hah1}). The current version of our interpolation code is not adapted to handle this kind of trail. We found less than five of these cases for the considered analysis set, so the impact vs effort ratio was deemed too low to add this feature to our code. This option should be considered for further work if larger data sets should be analysed.
  \item Trails presenting background or close sources modifying the extracted coordinates of the trail (see \citet{hah1} for the full trail coordinate extraction details) or other imaging artifacts such as trails slipping out of the cutouts. As mentioned before, our code uses both the starting and ending points as boundary conditions to obtain the object's rates of motion along with the image exposure time. It is specially sensitive to errors from incorrectly defining the trail's two ends. 
  \item Low signal to noise ratio trails, making coordinates difficult to obtain and trail limits difficult to measure.
\end{itemize}

\begin{figure}
   \centering
   \includegraphics[width=\columnwidth]{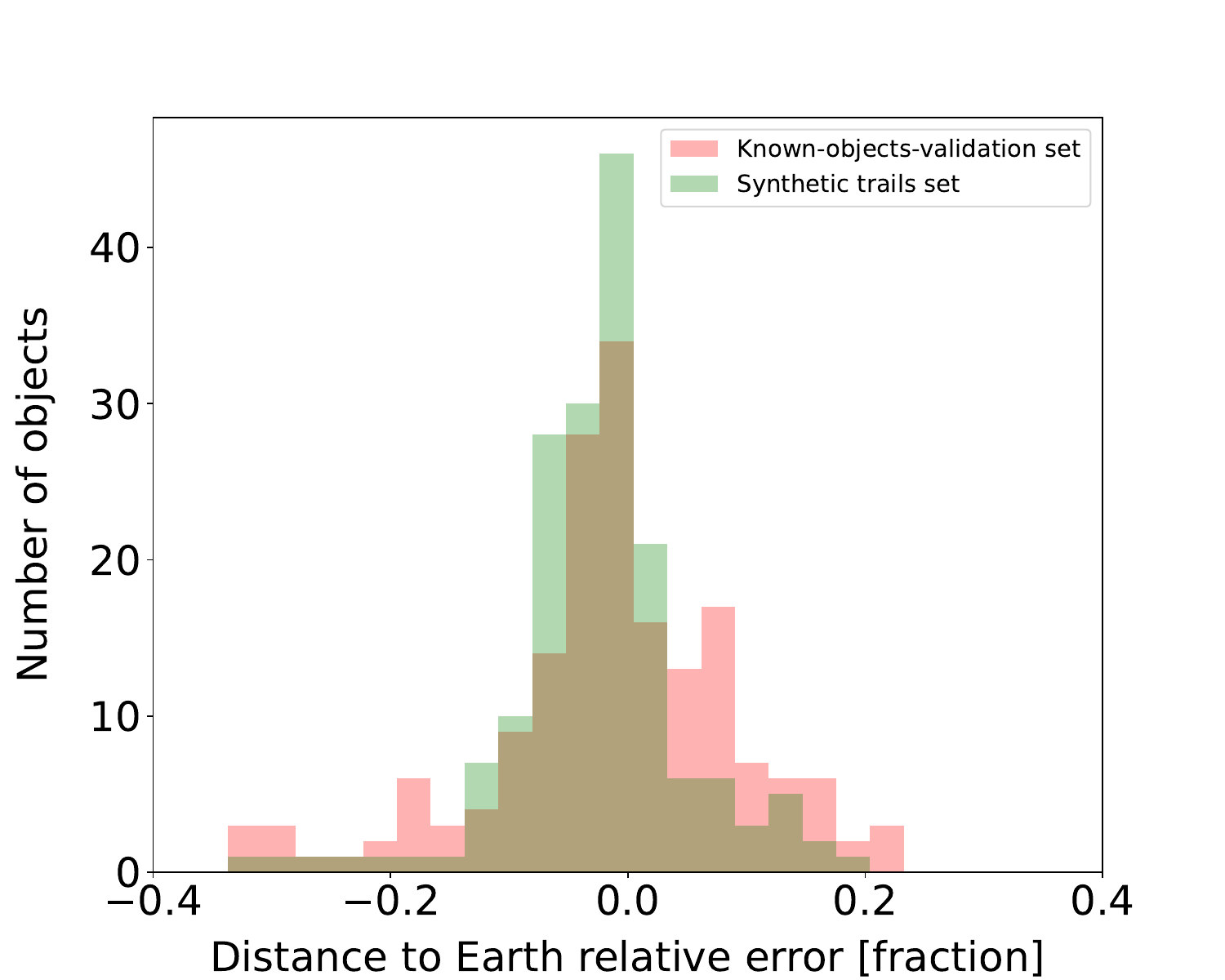}
   \caption{Distance to Earth relative error distribution [AU] for our known-objects-validation set using HST trails (red) compared to the relative error distribution for the synthetic trails validation set created using JPL Horizons ephemeris for these same objects and time (green).}
    \label{errors_histogram}%
\end{figure}

We applied these criteria keeping a total of 178 cases as our final known-objects-validation set. We consider these cases as a representative set to evaluate the accuracy of our parallax-based method, having removed any external influence from HST imaging or from our processing pipeline. 

This filtering process features a very conservative approach for our selection criteria which led us to keeping $\approx 40\% $ of the initial processed asteroids, optimising for detection quality over sample size. 

The distance to Earth relative error distribution for this known-objects-validation set is presented in Figure \ref{errors_histogram} ($ distance \ error/actual \ distance $). This error was calculated comparing JPL Horizons Earth distance with the computed distance using the parallax method. We obtain a mean relative error value centered at $-1.2\%$ and a standard deviation $\sigma = 10.5 \%$. This error distribution is presented to the reader as a global performance indicator of our novel parallax method. It will not be applied as an average error value to the results of this work. The specific distance uncertainty is calculated for each object as explained in Section \ref{parallax_section}.

To conduct a final verification of our method, we created a synthetic validation set. Starting from the known objects validation set, we obtained "synthetic" trails from JPL Horizon ephemeris for the same objects and exposure time as seen from HST. We fed these trails to our code as real HST trails to obtain a distance solution. In this case the obtained mean relative error value is $-2.5\%$ and $\sigma = 7.5\%$. The comparison of both error distributions is shown in Figure \ref{errors_histogram}. Both error distributions are closely overlapping and centered around 0, therefore we consider that any major influence from HST imaging or from our image processing pipeline was correctly removed from our analysis.

\begin{figure}
   \centering
   \includegraphics[width=\columnwidth]{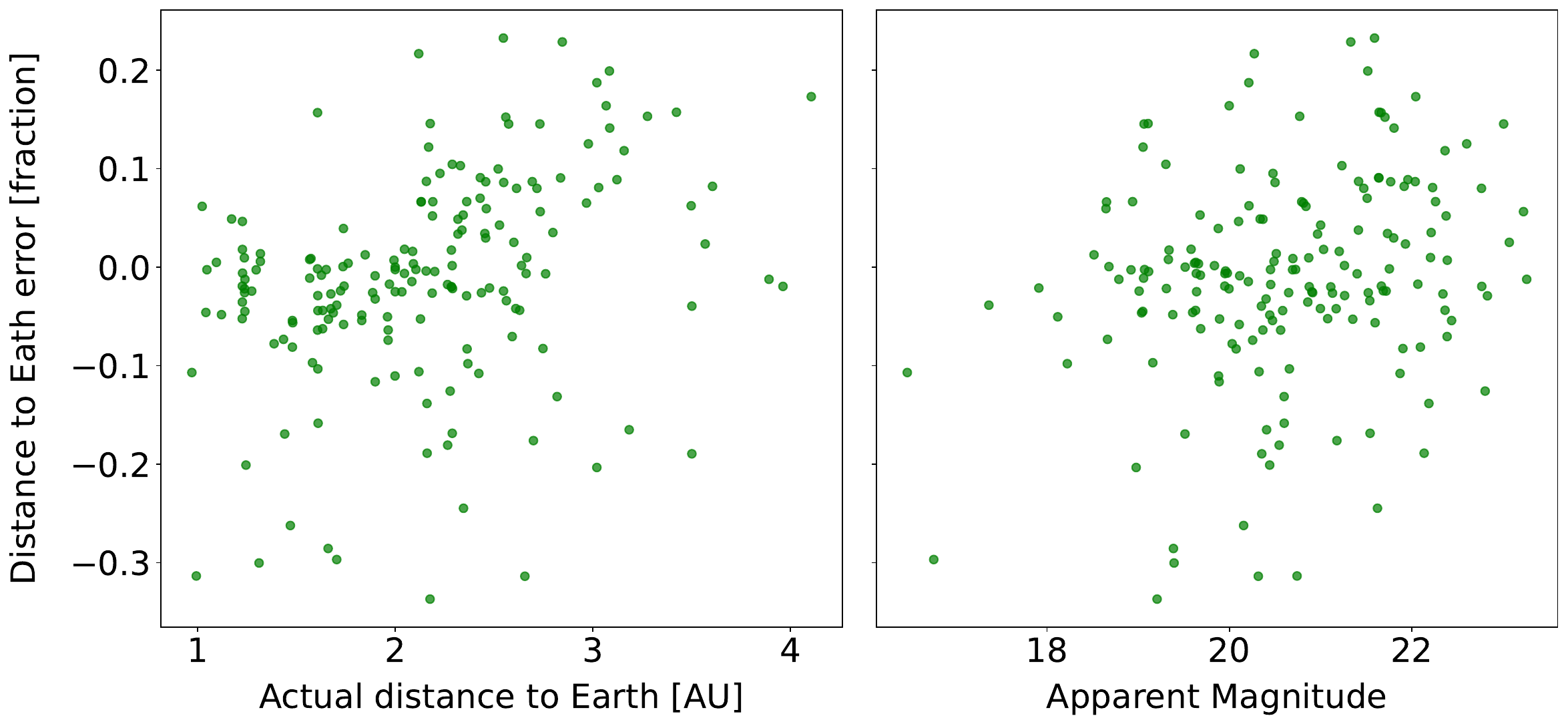}
   \caption{Scatter plot comparing the distance to Earth relative error obtained from the parallax method vs object's actual Earth distance and object's apparent magnitude}
    \label{errors_scattered}%
\end{figure}

As expected, we find a slight correlation between the object's actual distance to Earth or its apparent magnitude and the distance error using the parallax method (see Figure \ref{errors_scattered}). In general, far (and probably faint) objects yield more difficult trails to analyse due to lower signal to noise ratios and reduced parallax-induced curvature in the trails. 

\begin{figure}
   \centering
   \includegraphics[width=\columnwidth]{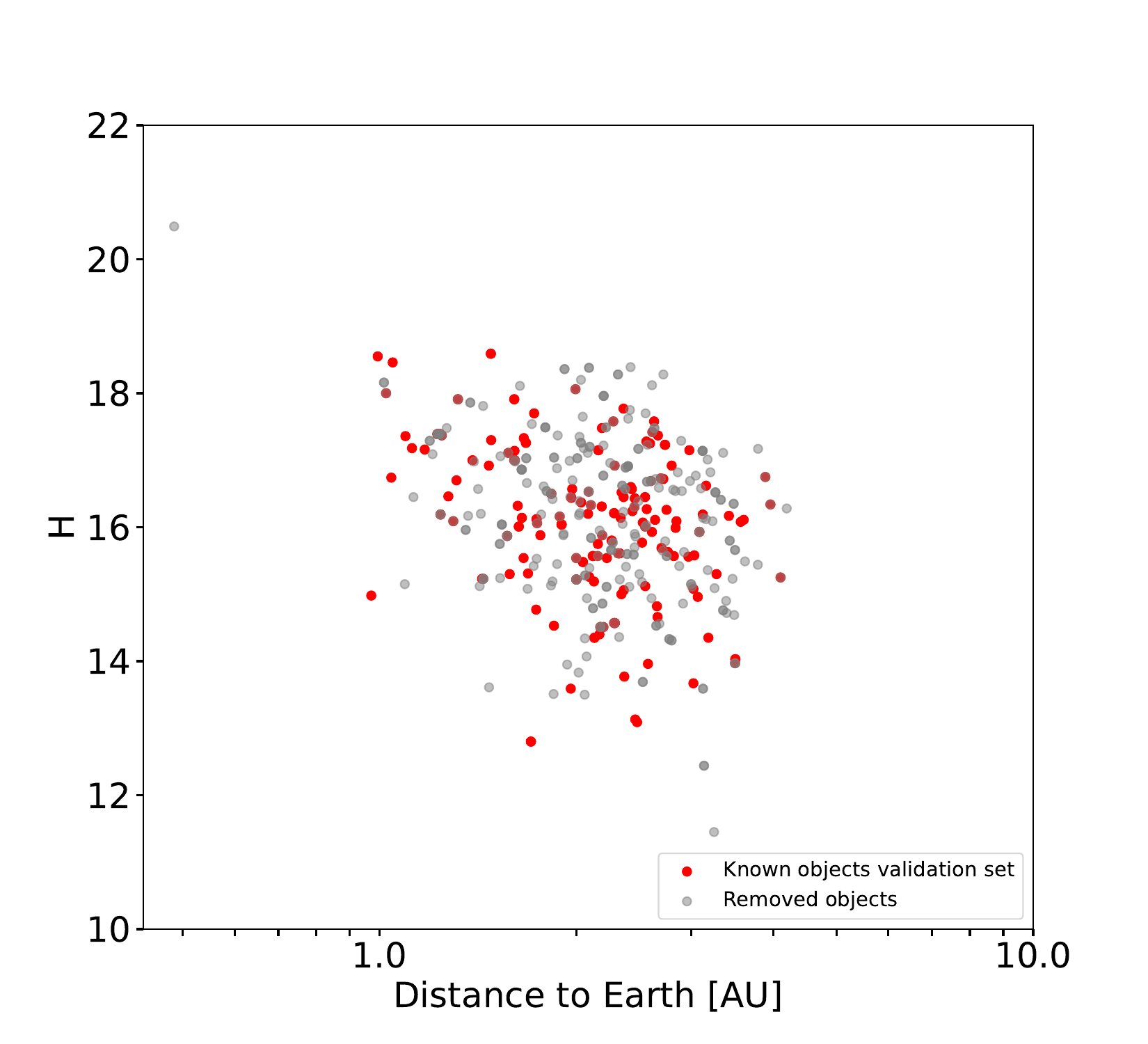}
   \caption{Absolute Magnitude vs Distance to Earth plot comparing our final known objects validation set (red) with those removed during the filtering process (grey). We can see a fairly random removal process around both axis, without any effect depending on distance or magnitude.}
    \label{removed_known}%
\end{figure}

Another important aspect to assess the accuracy of our method is to make sure the filtering process does not present any bias removing fainter or smaller objects, potentially reducing the number of outer Main Belt asteroids in our observations. We display in Figure \ref{removed_known} the initial sample of known objects (452 asteroids) as a function of distance to Earth and absolute magnitude (both data obtained from JPL Horizons, considered as ground truth). Two different populations are displayed on it: our final known objects validation set in red and the objects which were dropped during the filtering process in grey. We see a fairly constant and random removal process following both axis, without any systematic or selection effects. As intended, our filtering process is mainly driven by the trail characteristics, which are random, and not the physical properties of the asteroids themselves.

\begin{figure}
   \centering
   \includegraphics[width=\columnwidth]{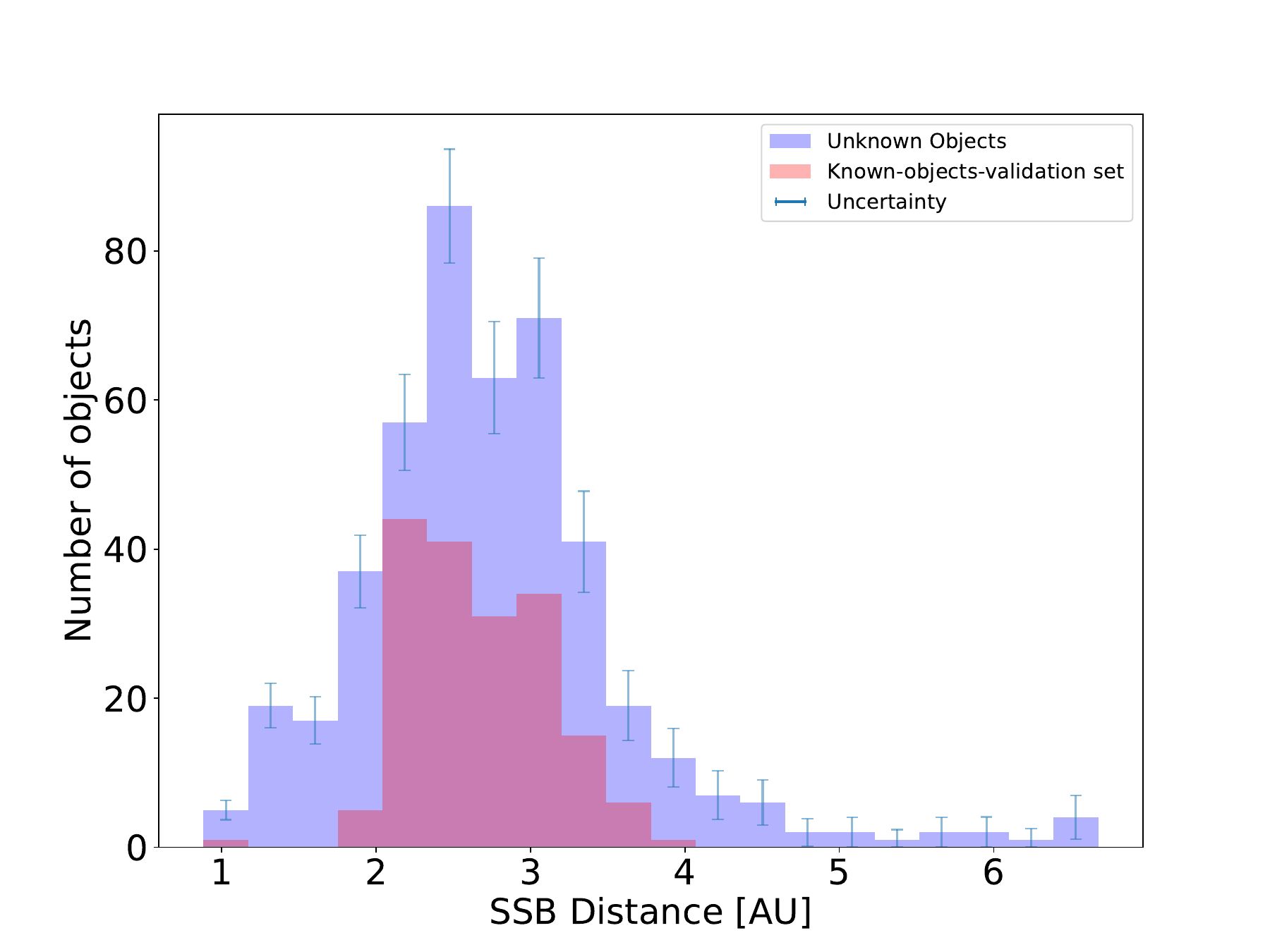}
   \caption{Distance to SSB for known (red) and unknown (blue) objects. Unknown object values were obtained from the parallax method, known object values from JPL Horizons database.}
    \label{distance_SSB}%
\end{figure}

\begin{figure}
   \centering
   \includegraphics[width=\columnwidth]{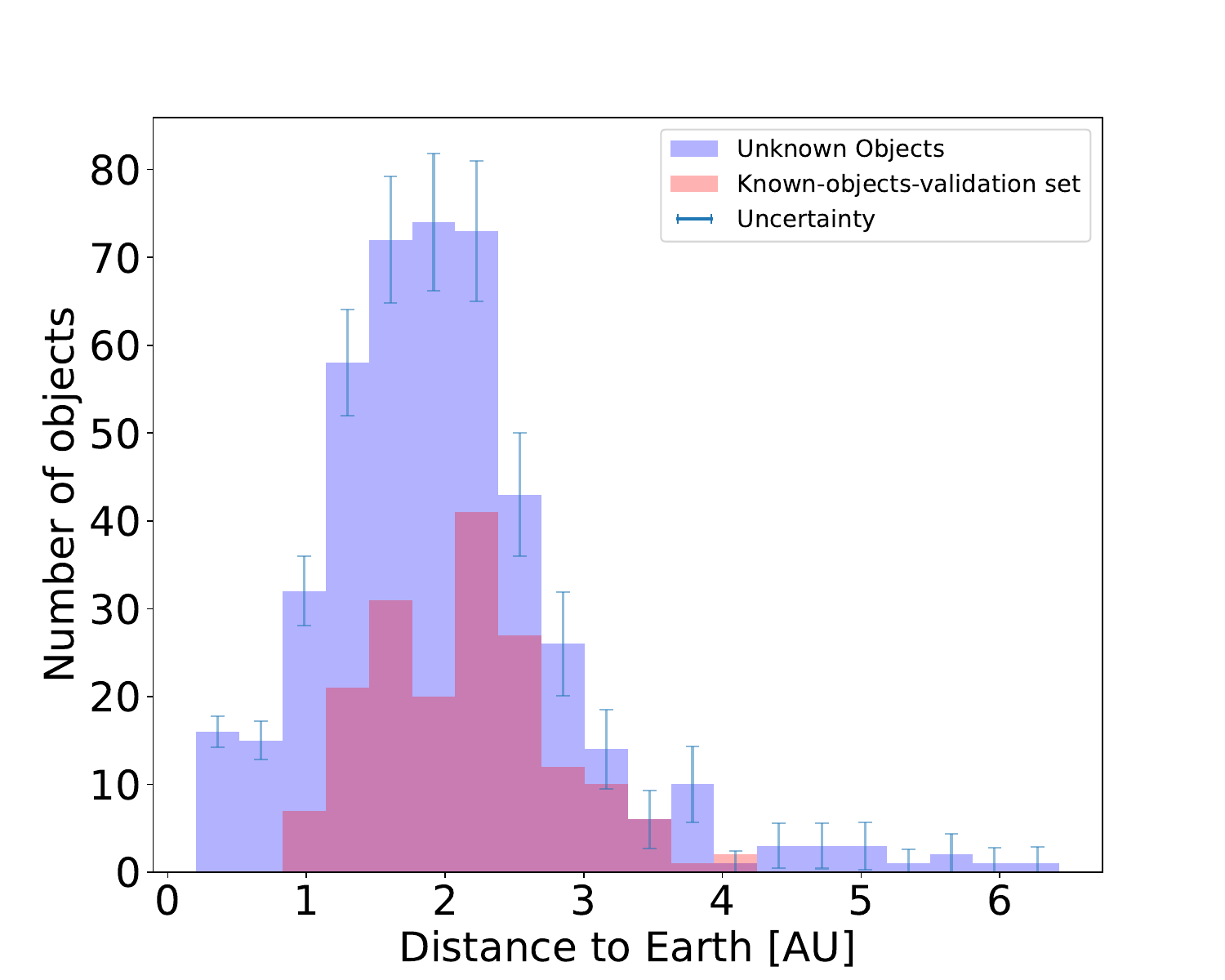}
   \caption{Distance to Earth for known (red) and unknown (blue) objects. Unknown object values were obtained from the parallax method, known object values from JPL Horizons database.}
    \label{distance_Earth}%
\end{figure}

\begin{figure}
   \centering
   \includegraphics[width=\columnwidth]{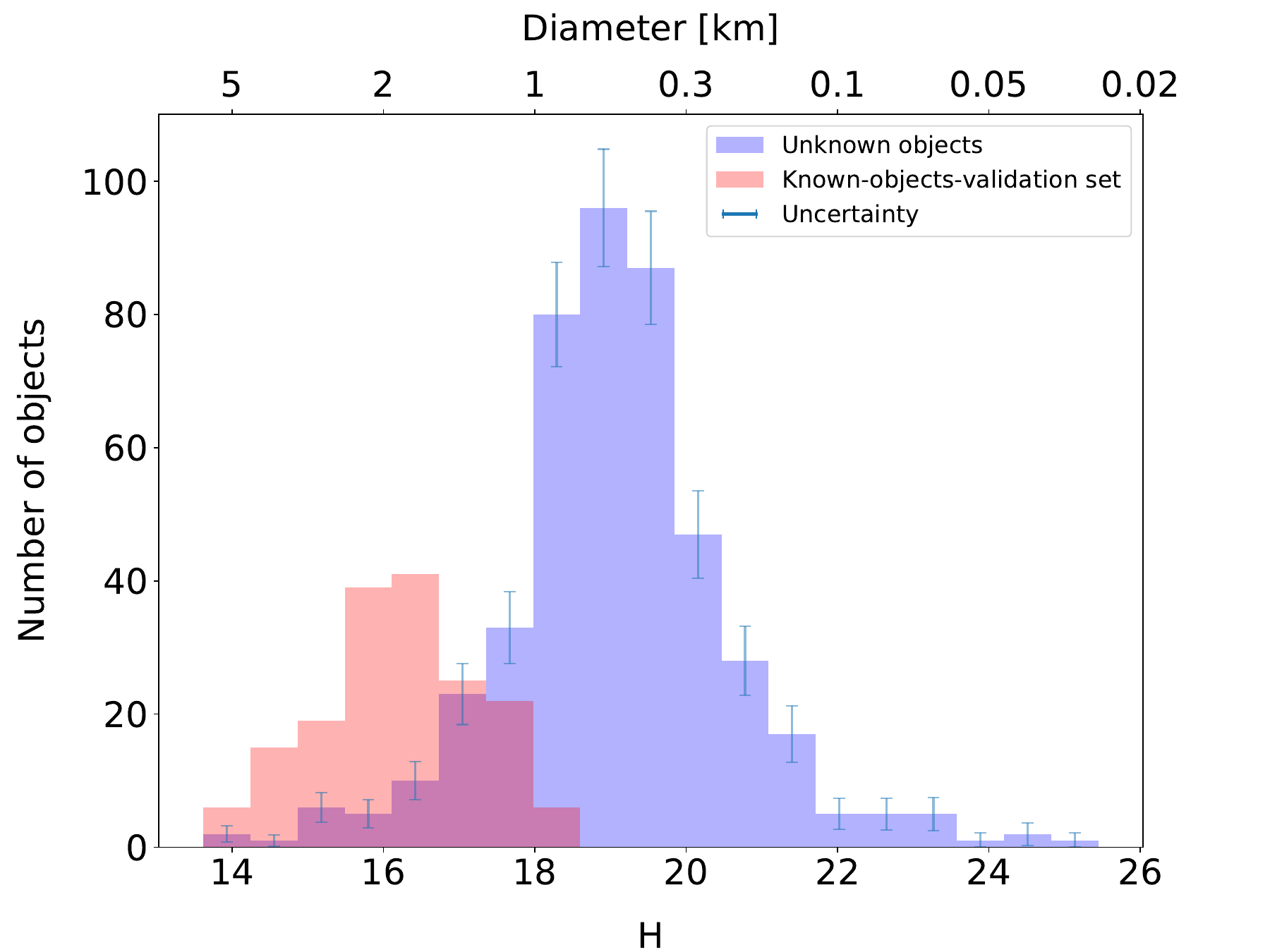}
   \caption{Absolute magnitude (H) for known (red) and unknown objects (blue) from JPL Horizons database and calculated using the distance obtained with the parallax method respectively. The upper axis shows the approximate object size using a 0.15 albedo value.}
    \label{H}%
\end{figure}

\subsection{Magnitude analysis}

The trails' apparent magnitudes were obtained in the previous part of this project \citep{hah1}. We start from these values to obtain the objects' absolute magnitudes using the distance solutions obtained with the parallax method.

Most of the unknown objects were serendipitiously imaged by HST in one single band, not having any further information about their color or asteroid type. Even if some of the our detections come from the same HST observation set, we cannot link the objects from one image to another, not having any reliable information about their orbital parameters. Before calculating their absolute magnitude values, we transform each object's apparent magnitude in any given HST filter to a Johnson V filter to apply an unified method to compare all of them. We assume the asteroids spectra to be mostly the reflected spectrum from the Sun, except for their particular absorption features, so the filter transformation value applied is the same used for the Sun spectrum. Using once again the known objects set as proof, we obtain a 0.18 mag average error for this approximation.

Using the objects's parallax-calculated distance and phase angle, we obtain the absolute magnitude using the equation from \citet{bowell}:

\begin{equation}
\begin{aligned}
H = V + 2.5log_{10}((1-G)\phi_1+G\phi_2) - 5log_{10}(\Delta \cdot r)
\end{aligned}
\label{eq:H}
\end{equation}

\noindent{where \textit{H} is the object's absolute magnitude, \textit{V} is its apparent magnitude in V, \textit{G} is the photometric slope parameter, \textit{$\Delta$} is object's distance to Earth [AU] and \textit{r} is object's distance to the Sun [AU]. We used a G=0.15 constant value for our analysis. \textit{$\phi_1$} and \textit{$\phi_2$} parameters are functions of the object's phase angle \(\alpha\).}
\\

\subsection{Size estimation method}
We use the absolute magnitude values (H) calculated in the previous section to estimate the size of the analysed objects, assuming them spherical and featuring a uniform surface and therefore constant albedo. To do so, we apply the equation from \citet{HARRIS1997450}. Even if albedo is known to change from S dominated to C dominated taxonomy classes from inner to outer belt \citep{DeMeoCarry}, an average 0.15 albedo value was chosen for estimating the object's size. This is an intermediate value representative for the typical range (0.03 - 0.4) of asteroid compositional types \citep[e.g.][]{2011ApJ...741...90M, 2018A&A...617A..12P, Mahlke22}. To avoid any error induced by this assumption, we will use 
 absolute magnitude (H) to create the population size distribution graph which will be discussed in Section \ref{discussion} (Figure \ref{population_H}). As an example, a H =19.5 object (around the median value for our unknown objects set) would feature a $\approx 700 \ m$ diameter supposing a 0.05 albedo and a $\approx 330\ m$ diameter if we suppose a 0.25 albedo.

\section{Results}
\label{results}

We apply the methods described to the set of 1,031 unknown Solar System Objects from \citet{hah1}. The strict filtering criteria we presented in Section \ref{accuracy} yield a total of 454 objects. This filtered set is $\approx 44\% $ of the original number, matching the yield obtained for the known-objects-validation set after cleansing. As stated in previous sections, we have prioritised high precision over a large population of objects. The majority of the input objects are purposely excluded during these filtering processes to maximise accuracy.

For the Results section histograms, data for known objects are extracted directly from JPL Horizons database. It is considered as baseline and therefore it does not feature any uncertainty bars in any of the presented graphs. The uncertainty bars for the unknown objects histograms are obtained using bootstrapping \citep{bootstrap}, randomly resampling the existing population 1,000 times and calculating the standard deviation for each bin.

\subsection{Object distance}

We obtained the distance to Earth and the distance to the Solar System Barycenter (SSB) for the 454 unknown objects. Results can be seen in Figures \ref{distance_SSB} and \ref{distance_Earth}. Both distance distributions feature a close overlap for known and unknown objects, centered around the Main belt in both cases. There is a small bump for unknown objects at less than 1 AU from Earth, representing close objects spotted by HST. Given the limited nature of our calculated orbital parameters, we cannot obtain a real perihelion value and therefore assess if they could be Near Earth Objects (NEO), as we will argue in Section \ref{discussion}.

\subsection{Magnitude and size distributions}

Despite having obtained a very close distance distribution for known and unknown objects (either to Earth or SSB), if we take a look to the absolute magnitude results displayed in Figure \ref{H} we can see that the unknown objects population is dominated by fainter objects, centered around H = 19-20 magnitude. Both results indicate that we are mostly imaging small (< 1 km) Main Belt objects, not easily accessible using ground based asteroid surveys.

\begin{figure}
   \centering
   \includegraphics[width=\columnwidth]{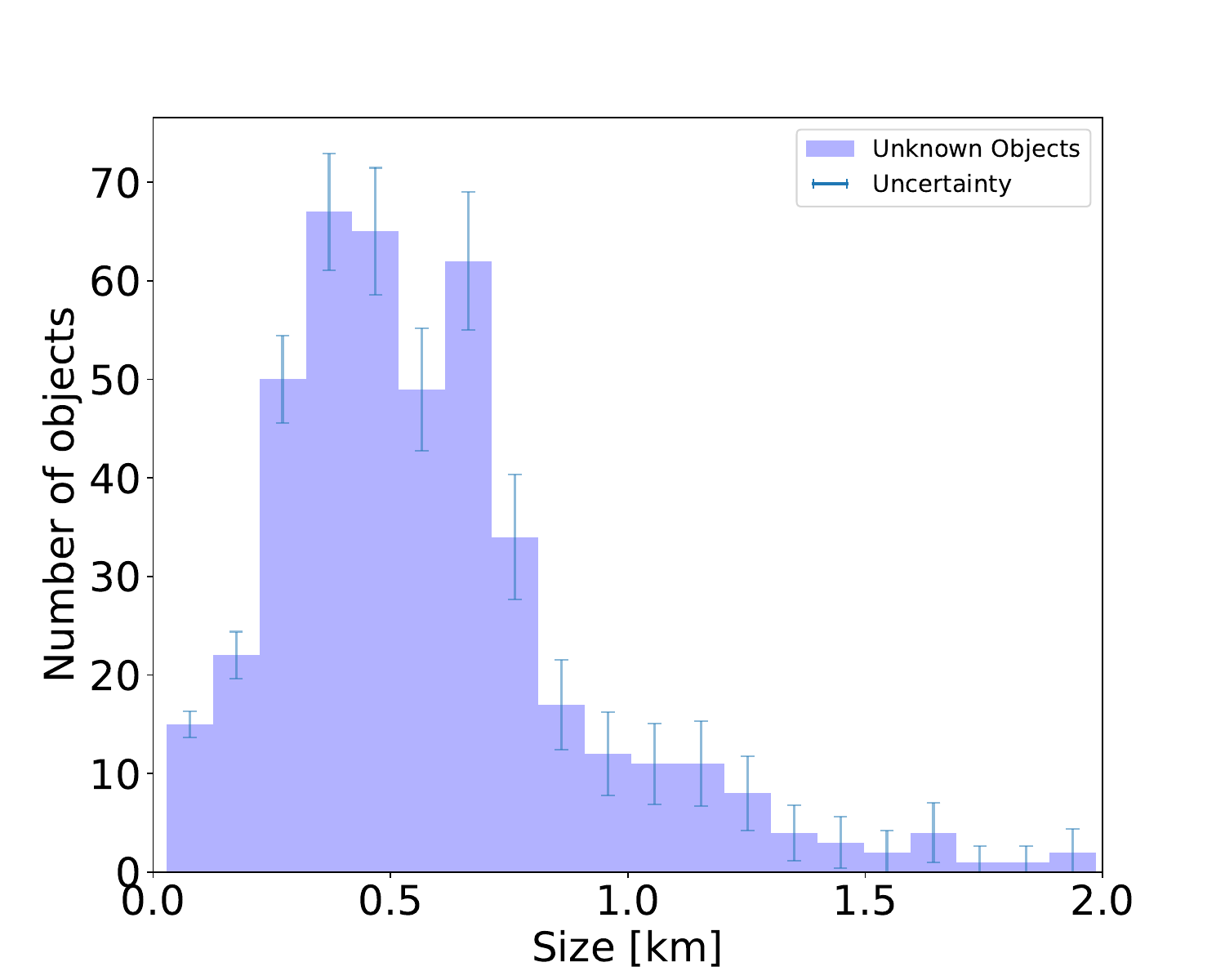}
   \caption{Size distribution for unknown objects (using a 0.15 albedo value).}
    \label{results_size_015}%
\end{figure}

The size distribution for the unknown objects (assuming an average albedo of 0.15 for all objects) is presented in Figure \ref{results_size_015}. A significant part of them are below 1 km in size. 

\begin{figure}
   \centering
   \includegraphics[width=\columnwidth]{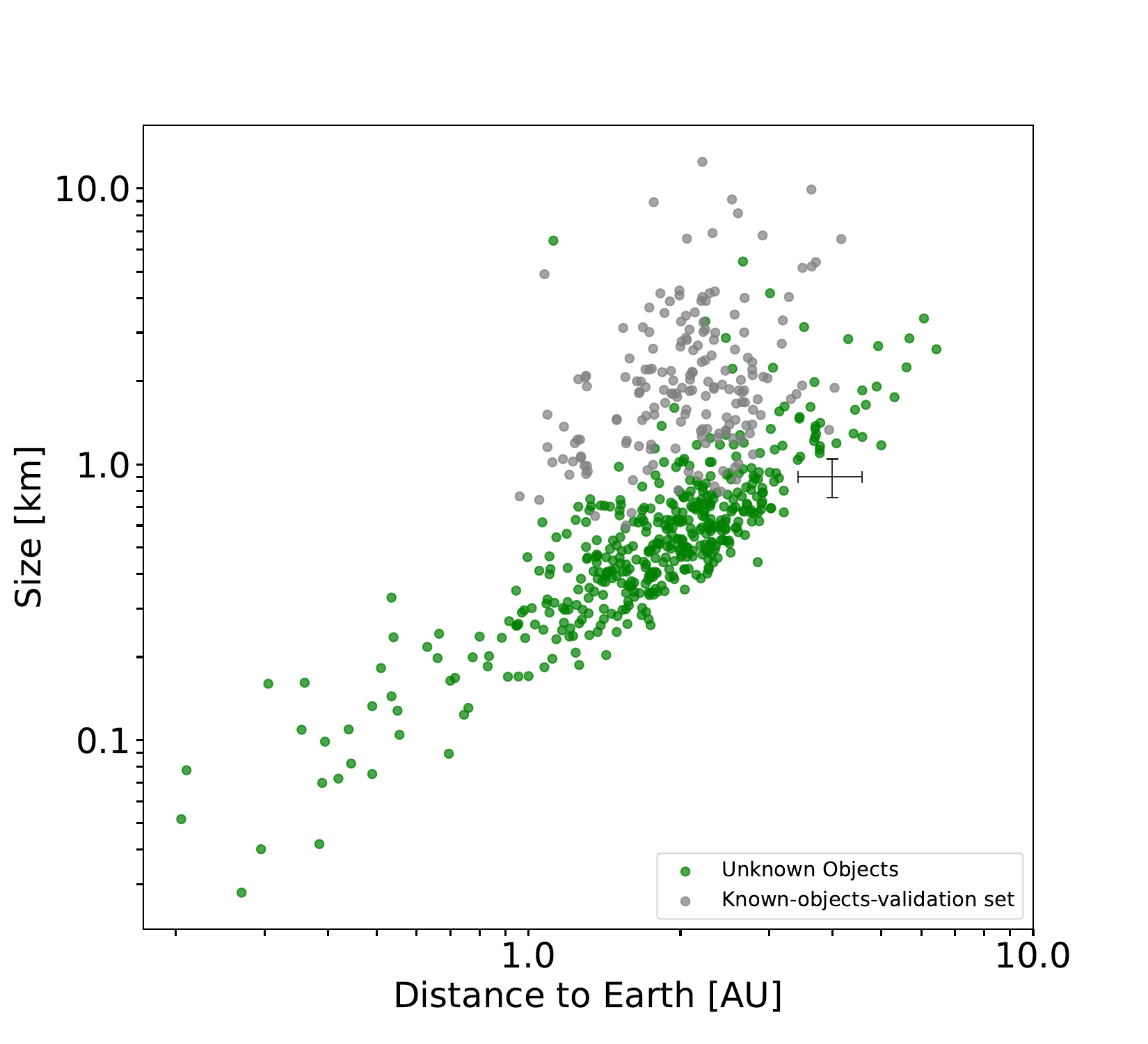}
   \caption{Size vs Earth distance for the known-objects-validation set (grey) and the unknown objects set (green, using a 0.15 albedo value to estimate size). For clarity purposes, the average size and distance uncertainties are displayed next to the main population area as a black cross.}
    \label{size_vs_distance}%
\end{figure}

\section{Discussion}
\label{discussion}

\subsection{Size and absolute magnitude population distributions}

A scatter plot presenting size as a function of Earth distance is shown in Figure \ref{size_vs_distance}. As we can see, the lower size limit of the detected unknown objects increases with distance to Earth, as we were expecting for a flux-limited survey. HST's unique capabilities allow us to detect unknown objects up to distances around 3 AU from Earth (with the main population centered around 2 AU) and below a 1 km estimated diameter. We find that our unknown objects population is mainly composed of small Main Belt asteroids ($<$ 1 km size).

Assuming that the size distribution is the same everywhere in the asteroid belt and even if our total dataset (unknown + known objects) does not represent a dedicated or observationally unbiased asteroid survey, we can use our HST set as a random asteroid population sample, given the large time span covered by our HST archival analysis (19 years) and the statistically random HST pointing locations \citep{hah1}. As shown in Figures \ref{distance_SSB} and \ref{size_vs_distance}, most of our objects are found in the Main Belt. If we take a closer look at Figure \ref{size_vs_distance}, we can clearly see the cutoff generated by the lower part of the scatter plot representing the observational limit of HST for a given distance. For a distance of 2.5 AU from Earth, roughly the end of the outer Main Belt, our size detection limit is 0.5 km, $H\approx 19.5$ using a 0.15 albedo. We assume our dataset to be complete up to this magnitude value regarding the Main Belt asteroid population.

\begin{figure}
   \centering
   \includegraphics[width=\columnwidth]{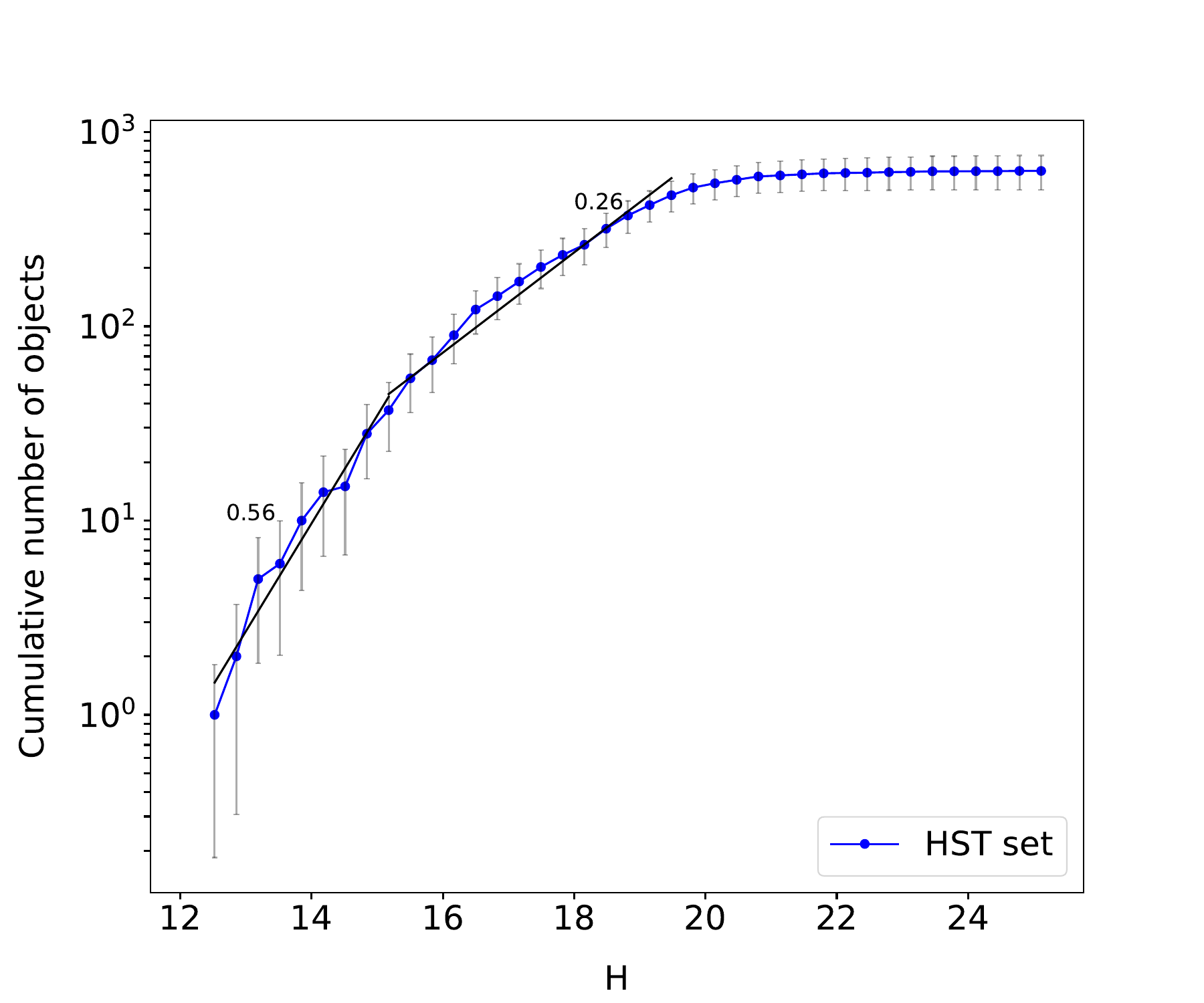}
   \caption{Absolute magnitude cumulative size distribution for our HST dataset. The black lines and numerical values represent the calculated the approximate equivalent slopes $\alpha$ for $logN(H>H_0)\propto\alpha\log(H_0)$. }
    \label{population_H}%
\end{figure}

The new population model we present is based on absolute magnitude (H) and is depicted in Figure \ref{population_H}. We present the entire HST set (unknown and known asteroids), featuring 632 objects. We identify two approximate different values for the slope of our cumulative distribution function: $log N(H>H_0)\propto(0.56 \pm 0.12)\log(H_0)$ for H values between 12 and 15, $log N(H>H_0)\propto(0.26 \pm 0.06)\log(H_0)$ for H values between 15 and 20. Above $H \approx 19-20 $, our dataset reaches its detection limit.

The values we obtained can be compared with the results from the Sub-Kilometer Asteroid Diameter Survey (SKADS) \citep{Gladman2009}. This dedicated and debiased survey for Main Belt small objects yielded a $\alpha = 0.3 \pm 0.02$ slope for H values between 15 and 18, in agreement to the value obtained in our analysis. The slight difference could be explained by our HST dataset bias or incompleteness and the different filters used in both analyses. The steeper slope change for the $H<15$ regime is also described in this work, with a reported $\alpha = 0.5$ value, very close to the one from HST asteroids set ($\alpha = 0.56$). Our work shows that the $\alpha = 0.3$ slope value between 15 and 18 H remains consistent up to a $H\approx 19-20$ mag. For reference, $H\approx15$ corresponds to a size of 3 km for a 0.15 albedo value. 

Other previous works have found results which are compatible with our survey. Using observations from WISE/NEOWISE in the thermal infrared, \citet{Masiero11} found a size distribution slope for Main Belt objects in good agreement with \citet{Gladman2009}, considered as the benchmark for our work. Still in the infrared and using data from Spitzer, \citet{Ryan15} found a slightly shallower population distribution slope with a "kink" or inflection point around a 8 km diameter, corresponding roughly to $H\approx13$ using a 0.15 albedo. In our case we do not have enough observations around this magnitude value to confirm this feature. \citet{Heinze19} performed Main Belt observations from the Blanco Telescope at Cerro Tololo Inter-American Observatory reporting a non-constant population power law slope. In this work, the apparent R-band population distribution shows a transition around R = 19-20, with a shallower slope for fainter magnitudes in good agreement with our benchmark \citep{Gladman2009} and maintained up to two fainter magnitudes than it, thus reinforcing the results from our work showing a constant slope maintained up to $H\approx 19-20$ mag. \citet{Maeda} found a similar shape for both "C-like" and "S-like" size population distributions featuring Main Belt asteroids between 0.4 and 5 km size (roughly $H\approx20$ and $H\approx14$ using a 0.15 albedo). This similarity between both types of asteroids supports the average nature of our population distribution, not being able to reliably distinguish colors in our unknown objects HST set. This work also reported an inflection point in the slope of the absolute magnitude population distribution, located around 1.7 mag fainter than our study. Slopes values are $\alpha = 0.55$ and $\alpha = 0.23$, slightly shallower for the fainter end.

\subsection{Potential comets}

An interesting possibility of our work is the existence of objects with orbits consistent to those of comets. To help us find possible candidates, we calculate the Tisserand's parameter \citep{1889BuAsI...6..241T} with respect to Jupiter ($T_J$) as seen in equation \ref{eq:TJ}, using the object's minimum inclination, eccentricity and semi-major axis values. 

\begin{equation}
\begin{aligned}
T_J = \frac{a_J}{a} + 2\sqrt{\frac{(1-e^2)a}{a_J}}\cdot cos(i)
\end{aligned}
\label{eq:TJ}
\end{equation}

\noindent{where \textit{$a_J$} is Jupiter's semi-major axis, \textit{a} is the object's minimum semi-major axis, \textit{e} is the object's minimum eccentricity and \textit{i} is the object's minimum inclination.}

\begin{figure}
   \centering
   \includegraphics[width=\columnwidth]{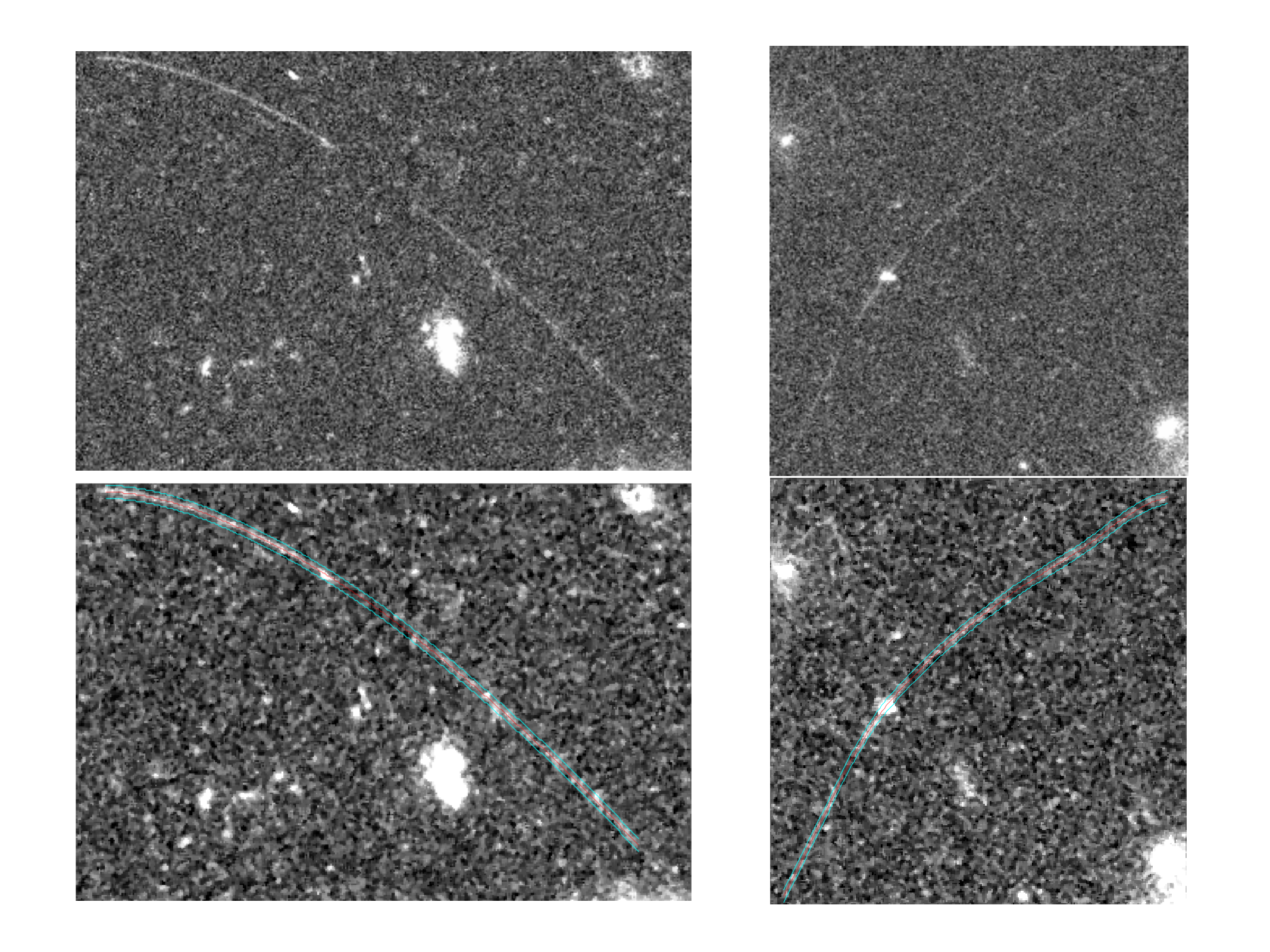}
   \caption{Two examples of potential comets found in out dataset: left images correspond to j8n218010 and right images to j90i02010. The upper row represents the original trail cutouts and the lower row the identified trail centerline and trail extraction lateral limits \citep{hah1}.}
   %\BC{add scale and orientation}}
    \label{potential_comets}%
\end{figure}

We find 45 objects of interest featuring $T_J$ < 3. This represents $\approx 10\%$ of the total 454 analysed objects. Two examples are displayed in Table \ref{comet_table} and Figure \ref{potential_comets}. All the 45 candidates present apparent magnitudes between 22.2 and 25.1 mag in the given filters used by HST for each observation, making them realistic candidates to new objects featuring comet orbits.

We note that $T_J$ parameter provides a first indication of a cometary orbit. The Jupiter family comets are characterized by $2 \leq T_J \geq 3$, while for the long period comets $T_J < 2$. An enhanced criterion to identify cometary orbits is provided by \citet{tancredi2014}. This takes into account the orbital parameters, the minimum orbital intersection distance and  discards the objects in mean motion resonances.

\begin{table}[]
    \centering
    \caption{Two examples of potential comets found in our dataset and their calculated approximate orbital parameters.}
    \begin{tabular}{  c | c | c  } 
    \hline\hline
    \textbf{Image} & \textbf{j8n218010} & \textbf{j90i02010}\\ 
    \hline
    Earth dist. [AU] & 0.55 +0.02 -0.03 & 2.91 +0.10 -0.10  \\ 
    % \hline
    SSB dist. [AU] & 1.52 +0.01 -0.03 & 3.10 +0.10 -0.10\\
    % \hline
    H & 22.15 +0.16 -0.08 & 18.35 +0.14 -0.14 \\
    % \hline
    $T_J$ (I min) & 1.69 & 1.07 \\
    % \hline
    e & 0.91 < e & 0.52 < e \\
    % \hline
    i [deg] &  17.35 < i < 17.63 & 82.08 < i < 99.79\\
    % \hline
    a [AU] & 15.59 < a & 6.44 < a\\
    % \hline
    Min. Perihelion [AU] & 1.37  & 3.08\\
    \hline
    \end{tabular}
    \label{comet_table}
\end{table}

 \subsection{Potential Near Earth Objects}

Near Earth objects (NEOs) are also a relevant class of objects that deserves a more detailed analysis to find potential candidates in our dataset. We define NEOs as objects featuring a perihelion < 1.3 AU \citep{BottkeNeos}. Considering that we just have minimum values for eccentricity and semi-major axis, the calculated perihelion and aphelion values are also approximations and they should be used just as an indicator of potential objects belonging to this family (shown in the lower-left corner of Figure \ref{size_vs_distance}). We find 74 potential NEO candidates. The apparent magnitudes in this case span from 20.4 to 25.1 in the given filter used by HST to take the image. 

Among our NEO set we find $\sim$ \textit{34\%} Amors, $\sim$ \textit{41\%} Apollos and $\sim$ \textit{12\%} Atens. These values are close but not coincident with the work from \citet{Granvik}, presenting a population distribution for NEOs in the range \textit{17 < H < 25} and featuring a relative share of \textit{39.4\%} Amors, \textit{54.4\%} Apollos and \textit{3.5\%} Atens. In our case, our analysis is limited by the approximate nature of the orbital parameters and HST observational bias for NEOs. Our dataset does not represent a complete population model for this kind of object (which is beyond the scope of this work) but a random sample of observations from HST archival data. 

\subsection{Further Work}
\label{further_work}

Absolute magnitude distributions can be used to estimate the size distribution, after a careful estimation of the albedo values, such as in the study conducted by \citet{Sloan} using commissioning data from SDSS and including color information. 

We are also currently analysing the lightcurves from all the detected asteroids (known and unknown objects). This analysis should lead to better characterisation of the properties of small objects, important to understand the collisional and dynamical Main Belt evolution \citep{Bottke_ast4}. The raw lightcurves from this project need specialised processing to take into account the parallax effect and correctly project time vs brightness variation. 

Beyond the discovery and physical characterisation of small asteroids in the HST archive, all the three papers from the Hubble Asteroid Hunter project illustrate how the training of deep learning neural networks to identify objects in images or other features in science data archives, using labels produced by large citizen science projects, could be used to automatically tag objects or phenomena in scientific datasets and assist data analysts later on the identification of such types of objects or phenomena, enlarging greatly the scientific potential of these datasets themselves. Encouraged by these prolific citizen science project, ESA has launched another two similar citizen science projects (Rosetta Zoo\footnote{\url{https://www.zooniverse.org/projects/ellenjj/rosetta-zoo/}} and GaiaVari\footnote{\url{https://www.zooniverse.org/projects/gaia-zooniverse/gaia-vari}}) where volunteers are asked to classify moving features in the surface of the comet 67P as observed by the Rosetta mission or classify different types of variable stars in the Gaia catalogue. The results of these projects will be also used to train deep neural networks that will allow the identification and classification of such features in future datasets (from similar or the same missions) reliably and automatically. In the particular case of the asteroids, serendipitous observations made by JWST will probably become a substantial source of new asteroid detections, as demonstrated recently in \citet{Mueller2023}. ESA's Euclid mission will also enable many new object discoveries in the years to come, thanks to its large field of view combined with its spatial resolution \citep{B_Euclid, Euclid_streak}. Unfortunately in both JWST and Euclid cases, their orbits around L2 are not suited to induce parallax like HST.

\section{Conclusions}
\label{conclusions}
We have analysed the physical properties of a set of 632 serendipitously imaged asteroids from HST archive. These asteroids were found using a combination of a Citizen Science project and a Machine Learning algorithm during previous work \citep{hah1}. To obtain asteroid distance from just one observation, we applied a proven method based on the parallax effect generated by HST's low Earth orbit \citep{Evans1998}. Approximately 40\% of the raw objects found in the archive yielded meaningful distance results using this method. We used these distance results to compute the asteroids absolute magnitude and approximate size. 

Our results indicate a set dominated by Main Belt objects featuring sizes below 10 km and absolute magnitudes (H) mostly between 15 and 22. The unknown objects set (454 objects) is bounded below 1 km size and between 17 and 22 for H. The absolute magnitude cumulative distribution confirms the previously reported slope change at H = 15 for Main Belt objects, from $\alpha \approx 0.56$ to $\alpha \approx 0.26$, \citep{Gladman2009}, maintained up to absolute magnitudes around $H\approx 20$ in our case. 

One of the advantages of applying machine learning to find Solar System objects in complete astronomical archives is the large number of potential results obtained. This allows us to apply purposely strict filtering conditions to improve accuracy and still keep a large enough sample to obtain statistically meaningful results.

Adding automatic SSO detection pipelines to deep but FoV-limited telescopes could generate important contributions to the population of small SSOs and to our understanding of the early stages and evolution of the Solar System. In our case, the analysis of archival data spanning 19 years limits the possibilities of any object follow-up for interesting cases such as potential NEOs or comets. This drawback can be avoided using real-time processing pipelines.

We have contacted the IAU Minor Planet Center to submit the full identifications set (known and unknown objects) for the whole Hubble Asteroid Hunter Project \citep{hah1}.

\begin{acknowledgements}
We thank the anonymous referee for the helpful comments and suggestions that led to the improvement of this manuscript. We acknowledge support from ESA through the Science Faculty - Funding reference ESA-SCI-SC-LE-154. This work was possible thanks to the Google Cloud Research Credits Program. The work of MP was supported by a grant of the Romanian National Authority for Scientific  Research -- UEFISCDI, project number PN-III-P2-2.1-PED-2021-3625. We want to specially thank the STScI helpdesk for their quick responses and useful advice. This publication uses data generated via the https://www.zooniverse.org/
platform, development of which is funded by generous support, including a
Global Impact Award from Google, and by a grant from the Alfred P. Sloan
Foundation. This work has made extensive use of data from HST
mission, hosted by the European Space Agency at the eHST archive, at ESAC
(https://www.cosmos.esa.int/hst), thanks to a partnership with the Space
Telescope Science Institute, in Baltimore, USA (https://www.stsci.edu/)
and with the Canadian Astronomical Data Centre, in Victoria, Canada
(https://www.cadc-ccda.hia-iha.nrc-cnrc.gc.ca/).

\end{acknowledgements}

\bibliographystyle{aa} % style aa.bst
\bibliography{bibliography.bib} % your references Yourfile.bib

\begin{thebibliography}{33}
\expandafter\ifx\csname natexlab\endcsname\relax\def\natexlab#1{#1}\fi

\bibitem[{{Acton} {et~al.}(2018){Acton}, {Bachman}, {Semenov}, \& {Wright}}]{spice2}
{Acton}, C., {Bachman}, N., {Semenov}, B., \& {Wright}, E. 2018, \planss, 150, 9

\bibitem[{{Acton}(1996)}]{spice}
{Acton}, C.~H. 1996, \planss, 44, 65

\bibitem[{{Berthier} {et~al.}(2016){Berthier}, {Carry}, {Vachier}, {Eggl}, \& {Santerne}}]{Skybot2}
{Berthier}, J., {Carry}, B., {Vachier}, F., {Eggl}, S., \& {Santerne}, A. 2016, \mnras, 458, 3394

\bibitem[{{Berthier} {et~al.}(2006){Berthier}, {Vachier}, {Thuillot}, {Fernique}, {Ochsenbein}, {Genova}, {Lainey}, \& {Arlot}}]{Skybot1}
{Berthier}, J., {Vachier}, F., {Thuillot}, W., {et~al.} 2006, in Astronomical Society of the Pacific Conference Series, Vol. 351, Astronomical Data Analysis Software and Systems XV, ed. C.~{Gabriel}, C.~{Arviset}, D.~{Ponz}, \& S.~{Enrique}, 367

\bibitem[{{Bottke} {et~al.}(2015){Bottke}, {Bro{\v{z}}}, {O'Brien}, {Campo Bagatin}, {Morbidelli}, \& {Marchi}}]{Bottke_ast4}
{Bottke}, W.~F., {Bro{\v{z}}}, M., {O'Brien}, D.~P., {et~al.} 2015, in Asteroids IV, 701--724

\bibitem[{{Bottke} {et~al.}(2002){Bottke}, {Morbidelli}, {Jedicke}, {Petit}, {Levison}, {Michel}, \& {Metcalfe}}]{BottkeNeos}
{Bottke}, W.~F., {Morbidelli}, A., {Jedicke}, R., {et~al.} 2002, \icarus, 156, 399

\bibitem[{{Bowell} {et~al.}(1989){Bowell}, {Hapke}, {Domingue}, {Lumme}, {Peltoniemi}, \& {Harris}}]{bowell}
{Bowell}, E., {Hapke}, B., {Domingue}, D., {et~al.} 1989, in Asteroids II, ed. R.~P. {Binzel}, T.~{Gehrels}, \& M.~S. {Matthews}, 524--556

\bibitem[{{Carry}(2018)}]{B_Euclid}
{Carry}, B. 2018, \aap, 609, A113

\bibitem[{{DeMeo} \& {Carry}(2014)}]{DeMeoCarry}
{DeMeo}, F.~E. \& {Carry}, B. 2014, \nat, 505, 629

\bibitem[{{Dressel}(2022)}]{WFChandbook}
{Dressel}, L. 2022, in WFC3 Instrument Handbook for Cycle 30 v. 14, Vol.~14, 14

\bibitem[{{Efron}(1982)}]{bootstrap}
{Efron}, B. 1982, {The Jackknife, the Bootstrap and other resampling plans}

\bibitem[{{Evans} \& {Stapelfeldt}(2002)}]{Evans2002}
{Evans}, R.~W. \& {Stapelfeldt}, K.~R. 2002, in ESA Special Publication, Vol. 500, Asteroids, Comets, and Meteors: ACM 2002, ed. B.~{Warmbein}, 509--512

\bibitem[{{Evans} {et~al.}(1998){Evans}, {Stapelfeldt}, {Peters}, {Trauger}, {Padgett}, {Ballester}, {Burrows}, {Clarke}, {Crisp}, {Gallagher}, {Griffiths}, {Grillmair}, {Hester}, {Hoessel}, {Holtzmann}, {Krist}, {McMaster}, {Meadows}, {Mould}, {Ostrander}, {Sahai}, {Scowen}, {Watson}, \& {Westphal}}]{Evans1998}
{Evans}, R.~W., {Stapelfeldt}, K.~R., {Peters}, D.~P., {et~al.} 1998, \icarus, 131, 261

\bibitem[{{Garvin} {et~al.}(2022){Garvin}, {Kruk}, {Cornen}, {Bhatawdekar}, {Ca{\~n}ameras}, \& {Mer{\'\i}n}}]{hah2}
{Garvin}, E.~O., {Kruk}, S., {Cornen}, C., {et~al.} 2022, \aap, 667, A141

\bibitem[{{Giorgini} {et~al.}(1996){Giorgini}, {Yeomans}, {Chamberlin}, {Chodas}, {Jacobson}, {Keesey}, {Lieske}, {Ostro}, {Standish}, \& {Wimberly}}]{jpl_horizons}
{Giorgini}, J.~D., {Yeomans}, D.~K., {Chamberlin}, A.~B., {et~al.} 1996, in AAS/Division for Planetary Sciences Meeting Abstracts, Vol.~28, AAS/Division for Planetary Sciences Meeting Abstracts \#28, 25.04

\bibitem[{{Gladman} {et~al.}(2009){Gladman}, {Davis}, {Neese}, {Jedicke}, {Williams}, {Kavelaars}, {Petit}, {Scholl}, {Holman}, {Warrington}, {Esquerdo}, \& {Tricarico}}]{Gladman2009}
{Gladman}, B.~J., {Davis}, D.~R., {Neese}, C., {et~al.} 2009, \icarus, 202, 104

\bibitem[{{Granvik} {et~al.}(2018){Granvik}, {Morbidelli}, {Jedicke}, {Bolin}, {Bottke}, {Beshore}, {Vokrouhlick{\'y}}, {Nesvorn{\'y}}, \& {Michel}}]{Granvik}
{Granvik}, M., {Morbidelli}, A., {Jedicke}, R., {et~al.} 2018, \icarus, 312, 181

\bibitem[{Harris \& Harris(1997)}]{HARRIS1997450}
Harris, A.~W. \& Harris, A.~W. 1997, Icarus, 126, 450

\bibitem[{{Heinze} {et~al.}(2019){Heinze}, {Trollo}, \& {Metchev}}]{Heinze19}
{Heinze}, A.~N., {Trollo}, J., \& {Metchev}, S. 2019, \aj, 158, 232

\bibitem[{{Ivezi{\'c}} {et~al.}(2001){Ivezi{\'c}}, {Tabachnik}, {Rafikov}, {Lupton}, {Quinn}, {Hammergren}, {Eyer}, {Chu}, {Armstrong}, {Fan}, {Finlator}, {Geballe}, {Gunn}, {Hennessy}, {Knapp}, {Leggett}, {Munn}, {Pier}, {Rockosi}, {Schneider}, {Strauss}, {Yanny}, {Brinkmann}, {Csabai}, {Hindsley}, {Kent}, {Lamb}, {Margon}, {McKay}, {Smith}, {Waddel}, {York}, \& {SDSS Collaboration}}]{Sloan}
{Ivezi{\'c}}, {\v{Z}}., {Tabachnik}, S., {Rafikov}, R., {et~al.} 2001, \aj, 122, 2749

\bibitem[{{Kruk} {et~al.}(2023){Kruk}, {Garc{\'\i}a-Mart{\'\i}n}, {Popescu}, {Aussel}, {Dillmann}, {Perks}, {Lund}, {Mer{\'\i}n}, {Thomson}, {Karadag}, \& {McCaughrean}}]{natastron_sats}
{Kruk}, S., {Garc{\'\i}a-Mart{\'\i}n}, P., {Popescu}, M., {et~al.} 2023, Nature Astronomy

\bibitem[{{Kruk} {et~al.}(2022){Kruk}, {Garc{\'\i}a Mart{\'\i}n}, {Popescu}, {Mer{\'\i}n}, {Mahlke}, {Carry}, {Thomson}, {Karada{\u{g}}}, {Dur{\'a}n}, {Racero}, {Giordano}, {Baines}, {de Marchi}, \& {Laureijs}}]{hah1}
{Kruk}, S., {Garc{\'\i}a Mart{\'\i}n}, P., {Popescu}, M., {et~al.} 2022, \aap, 661, A85

\bibitem[{{Maeda} {et~al.}(2021){Maeda}, {Terai}, {Ohtsuki}, {Yoshida}, {Ishihara}, \& {Deyama}}]{Maeda}
{Maeda}, N., {Terai}, T., {Ohtsuki}, K., {et~al.} 2021, \aj, 162, 280

\bibitem[{{Mahlke} {et~al.}(2022){Mahlke}, {Carry}, \& {Mattei}}]{Mahlke22}
{Mahlke}, M., {Carry}, B., \& {Mattei}, P.~A. 2022, \aap, 665, A26

\bibitem[{{Mainzer} {et~al.}(2011){Mainzer}, {Grav}, {Masiero}, {Hand}, {Bauer}, {Tholen}, {McMillan}, {Spahr}, {Cutri}, {Wright}, {Watkins}, {Mo}, \& {Maleszewski}}]{2011ApJ...741...90M}
{Mainzer}, A., {Grav}, T., {Masiero}, J., {et~al.} 2011, \apj, 741, 90

\bibitem[{{Masiero} {et~al.}(2011){Masiero}, {Mainzer}, {Grav}, {Bauer}, {Cutri}, {Dailey}, {Eisenhardt}, {McMillan}, {Spahr}, {Skrutskie}, {Tholen}, {Walker}, {Wright}, {DeBaun}, {Elsbury}, {Gautier}, {Gomillion}, \& {Wilkins}}]{Masiero11}
{Masiero}, J.~R., {Mainzer}, A.~K., {Grav}, T., {et~al.} 2011, \apj, 741, 68

\bibitem[{{M{\"u}ller} {et~al.}(2023){M{\"u}ller}, {Micheli}, {Santana-Ros}, {Bartczak}, {Oszkiewicz}, \& {Kruk}}]{Mueller2023}
{M{\"u}ller}, T.~G., {Micheli}, M., {Santana-Ros}, T., {et~al.} 2023, \aap, 670, A53

\bibitem[{{P{\"o}ntinen} {et~al.}(2020){P{\"o}ntinen}, {Granvik}, {Nucita}, {Conversi}, {Altieri}, {Auricchio}, {Bodendorf}, {Bonino}, {Brescia}, {Capobianco}, {Carretero}, {Carry}, {Castellano}, {Cledassou}, {Congedo}, {Corcione}, {Cropper}, {Dusini}, {Frailis}, {Franceschi}, {Fumana}, {Garilli}, {Grupp}, {Hormuth}, {Israel}, {Jahnke}, {Kermiche}, {Kitching}, {Kohley}, {Kubik}, {Kunz}, {Laureijs}, {Lilje}, {Lloro}, {Maiorano}, {Marggraf}, {Massey}, {Meneghetti}, {Meylan}, {Moscardini}, {Padilla}, {Paltani}, {Pasian}, {Pires}, {Polenta}, {Raison}, {Roncarelli}, {Rossetti}, {Saglia}, {Schneider}, {Secroun}, {Serrano}, {Sirri}, {Taylor}, {Tereno}, {Toledo-Moreo}, {Valenziano}, {Wang}, {Wetzstein}, \& {Zoubian}}]{Euclid_streak}
{P{\"o}ntinen}, M., {Granvik}, M., {Nucita}, A.~A., {et~al.} 2020, \aap, 644, A35

\bibitem[{{Popescu} {et~al.}(2018){Popescu}, {Licandro}, {Carvano}, {Stoicescu}, {de Le{\'o}n}, {Morate}, {Boac{\u{a}}}, \& {Cristescu}}]{2018A&A...617A..12P}
{Popescu}, M., {Licandro}, J., {Carvano}, J.~M., {et~al.} 2018, \aap, 617, A12

\bibitem[{{Ryan} {et~al.}(2015){Ryan}, {Mizuno}, {Shenoy}, {Woodward}, {Carey}, {Noriega-Crespo}, {Kraemer}, \& {Price}}]{Ryan15}
{Ryan}, E.~L., {Mizuno}, D.~R., {Shenoy}, S.~S., {et~al.} 2015, \aap, 578, A42

\bibitem[{{Ryon}(2022)}]{ACShandbook}
{Ryon}, J.~E. 2022, in ACS Instrument Handbook for Cycle 30 v. 21.0, Vol.~21, 21

\bibitem[{{Tancredi}(2014)}]{tancredi2014}
{Tancredi}, G. 2014, \icarus, 234, 66

\bibitem[{{Tisserand}(1889)}]{1889BuAsI...6..241T}
{Tisserand}, F. 1889, Bulletin Astronomique, Serie I, 6, 241

\end{thebibliography}

%\appendix
% \clearpage
% \newpage
% \onecolumn
% \LTcapwidth=\textwidth

\end{document}